\documentclass[journal,onecolumn,10pt,twoside]{IEEEtranTCOM}
\newtheorem{definition}{{Definition}}
\newtheorem{theorem}{{Theorem}}

\newtheorem{remark}{{Remark}}

\usepackage{cite}
\usepackage{amsmath,amssymb,amsfonts}
\usepackage{algorithmic}
\usepackage{graphicx}
\usepackage{textcomp}
\usepackage{xcolor}

\normalsize

\ifCLASSINFOpdf
\else
\fi

\begin{document}

\title{Channel-Aware Optimal Transport: A Theoretical Framework for Generative Communication}

\author{Xiqiang~Qu, Ruibin~Li,        Jun~Chen, Lei~Yu,
	and~Xinbing~Wang
}


\maketitle

\begin{abstract}
	Optimal transport has numerous applications, particularly in machine learning tasks involving generative models. In practice, the transportation process often encounters an information bottleneck, typically arising from the conversion of a communication channel into a rate-limited bit pipeline using error correction codes. While this conversion enables a channel-oblivious approach to optimal transport, it fails to fully exploit the available degrees of freedom. Motivated by the emerging paradigm of generative communication, this paper examines the problem of channel-aware optimal transport, where a block of i.i.d. random variables is transmitted through a memoryless channel to generate another block of i.i.d. random variables with a prescribed marginal distribution such that the end-to-end distortion is minimized. With unlimited common randomness available to the encoder and decoder,  the source-channel separation architecture is shown to be asymptotically optimal as the blocklength approaches infinity. On the other hand, in the absence of common randomness, the source-channel separation architecture is generally suboptimal. For this scenario, a hybrid coding scheme is proposed, which partially retains the generative capabilities of the given channel while enabling reliable transmission of digital information. It is demonstrated that the proposed hybrid coding scheme can outperform both  separation-based and uncoded schemes.
	
\end{abstract}

\begin{IEEEkeywords}
	Common randomness, generative model, hybrid coding, joint source-channel coding,  optimal transport, perception constraint, separation-based architecture, total variation distance, uncoded scheme, Wasserstein distance.   
\end{IEEEkeywords}

%
\IEEEpeerreviewmaketitle

\section{Introduction}

\IEEEPARstart{T}{he}  problem of optimal transport \cite{Villani03} can be mathematically formulated as reconstructing a random variable/vector with a prescribed distribution at the destination while minimizing the distortion relative to a given random variable/vector at the source. It has found numerous applications in fields such as economics, computer vision, machine learning, and data science, to name a few.

In practical scenarios, an information bottleneck often exists between the source and destination. One example is rate-limited optimal transport\footnote{It is also referred to as output-constrained lossy source coding.} \cite{LKK10,KZLK13,SLY15J1,SLY15J2,LZCK22,	LZCK22J,GPC24}, where the bottleneck is imposed in the form of a rate-constrained bit pipeline. This constraint restricts the set of feasible transportation plans that can be implemented. A notable variant of rate-limited optimal transport, known as rate-distortion-perception coding \cite{BM18, BM19, Matsumoto18, Matsumoto19, YWYML21,TW21, TA21, ZQCK21,CYWSGT22,Wagner22,SPCYK23,SCKY24,QSCKYSGT24,XLCZ24,XLCYZ24}, has garnered significant attention in recent years due to its potential as a framework for perception-aware image/video compression. In this problem, the reconstruction distribution is not fixed but is required to approximate the source distribution under a specified divergence, introducing flexibility and aligning the coding process with perceptual quality considerations.

In this paper, we study a joint source-channel coding version of the information-constrained optimal transport problem, where a communication channel serves as the information bottleneck. Unlike the aforementioned source-coding counterpart with a hard bit constraint, the bottleneck imposed by a communication channel is inherently softer and considerably more flexible. This added flexibility introduces greater degrees of freedom for exploration but also poses significant analytical challenges.

Our formulation is partly inspired by the surge of recent research on deep joint source-channel coding for image/video transmission \cite{BKG19,KG20,KG21,TG22,ETDG23}, where both the transmitter and receiver are implemented as trainable deep neural networks. While experimental results in this area have been highly promising, the theoretical foundation remains largely undeveloped. In particular, the employment of generative models and the shift way from the conventional distortion-oriented paradigm pose many new problems that do not directly fit into  Shannon's classical framework. 
We hope the present work contributes meaningfully to addressing this gap.

The main results of this paper are as follows. It is demonstrated that for the problem of channel-aware optimal transport, namely, transporting a block of i.i.d. random variables through a memoryless channel to produce another block of i.i.d. random variables with a prescribed marginal distribution while minimizing the end-to-end distortion, the source-channel separation architecture is asymptotically optimal as the blocklength approaches infinity, provided the encoder and decoder have access to unlimited common randomness. On the other hand, in the absence of common randomness, the source-channel separation architecture is shown to be generally suboptimal. Moreover, a hybrid coding scheme is proposed, which combines and enhances the strengths of both  separation-based and uncoded  schemes.

It is well-known that for the classical source-channel communication problem, joint coding can be strictly more powerful than separation-based coding in multiterminal settings \cite{CEGS80,TCDS14,SCT15,KC15,KC16,TCDS17}. However, in the point-to-point setting, the advantage of joint coding over separation-based coding vanishes in the asymptotic long blocklength limit, as established by Shannon's source-channel separation theorem \cite[Theorem 3.7]{EGK11}.
An important finding of this paper is that for the problem of channel-aware optimal transport, in the absence of common randomness, the separation architecture is strictly suboptimal even in the point-to-point setting. This suggests that the observed performance improvement of deep joint source-channel coding over separation-based coding cannot be solely attributed to the advantage of joint source-channel coding in the finite-blocklength regime---particularly in scenarios where a perception constraint is imposed on the reconstruction. A distinguishing feature of the problem under consideration, compared to the classical source-channel communication problem, is the generative aspect of  reconstruction. The key insight here is that instead of employing error correction codes to suppress the generative capabilities of communication channels, converting them into error-free bit pipelines, it is more advantageous to harness this generative power to reduce the decoder's burden. Since both channel randomness and generative decoding contribute to the end-to-end distortion, integrating these two sources of stochasticity can effectively improve the overall performance.



The rest of this paper is organized as follows. We introduce the problem definitions in Section \ref{sec:problemdefinition} and present the main results in Section \ref{sec:mainresults}. Sections \ref{sec:binarycase} and \ref{sec:Gaussiancase} provide in-depth analyses of the binary case and the Gaussian case, respectively. 
We conclude the paper in Section \ref{sec:conclusion}.

Notation: Let $\mathcal{B}(\alpha)$  represent the Bernoulli distribution with parameter $\alpha$, and $\mathcal{N}(\mu,\Sigma)$ represent the Gaussian distribution with mean $\mu$ and covariance matrix $\Sigma$. 
The $\delta$-typical set with respect to a given distribution is denoted by $\mathcal{T}^{(n)}_{\delta}$. For any real numbers $a$ and $b$, we abbreviate $\min\{a,b\}$, $(1-a)b+a(1-b)$, and $\max\{a,0\}$ as $a\wedge b$, $a*b$, and $(a)_+$, respectively. The trace operator is denoted by $\mathrm{tr}(\cdot)$, and the ceiling operatior is denoted by $\lceil\cdot\rceil$.  The indicator function is written as $1\{\cdot\}$.
Throughout this paper,  the logarithm is taken to base $2$.

\section{Problem Definitions}\label{sec:problemdefinition}

Let $\{X(t)\}_{t=1}^{n}$ be a block of i.i.d. random variables with marginal distribution $p_X$ over  alphabet $\mathcal{X}$. We aim to  construct another block of i.i.d. random variables $\{Y(t)\}_{t=1}^{n}$ with a prescribed marginal distribution $p_Y$ over  alphabet $\mathcal{Y}$ such that  
the (average) distortion $\frac{1}{n}\sum_{t=1}^n\mathbb{E}[d(X(t),Y(t))]$
is minimized, where $d:\mathcal{X}\times\mathcal{Y}\rightarrow[0,\infty)$ is a given distortion measure. Clearly, the minimum achievable distortion is
\begin{align}
	D^*:=\min\limits_{p_{XY}\in\Pi(p_X,p_Y)}\mathbb{E}[d(X,Y)],\label{eq:OT}
\end{align}
where $\Pi(p_X,p_Y)$ denotes the set of joint distributions with marginals $p_X$ and $p_Y$. Note that \eqref{eq:OT} corresponds to the standard optimal transport problem with $d(\cdot,\cdot)$ interpreted as the transportation cost function.

In this paper, we consider a scenario where $\{X_t\}_{t=1}^{n}$ and $\{Y_t\}_{t=1}^{n}$ are not co-located, and  transporation must occur through a memoryless channel $p_{V|U}$ with input alphabet $\mathcal{U}$ and  output alphabet $\mathcal{V}$. The presence of the channel imposes constraints on the set of feasbile transportation plans. Consequently, $D^*$ in \eqref{eq:OT} may no longer be achievable.
To determine the minimum achievable distortion in this scenario, we consider two different cases. 
\begin{enumerate}
	\item With common randomness:  In this case, the encoder and decoder share a random seed $Q$, and their operations are specified by  conditional distributions $p_{U^n|X^nQ}$ and $p_{Y^n|V^nQ}$, respectively. The entire system is characterized by joint distribution $p_{X^nQU^nV^nY^n}$, which factors as
	\begin{align}
		p_{X^nQU^nV^nY^n}=p_{X^n}p_Qp_{U^n|X^nQ}p_{V^n|U^n}p_{Y^n|V^nQ}\label{eq:factor1}
	\end{align}
with $p_{X^n}=p^n_X$ and $p_{V^n|U^n}=p^n_{V|U}$.
	
	\item Without common randomness: In this case, the encoder and decoder do not share a random seed, and their operations are specified by  conditional distributions $p_{U^n|X^n}$ and $p_{Y^n|V^n}$, respectively. The entire system is characterized by joint distribution $p_{X^nU^nV^nY^n}$, which factors as
	\begin{align}
		p_{X^nU^nV^nY^n}=p_{X^n}p_{U^n|X^n}p_{V^n|U^n}p_{Y^n|V^n}\label{eq:factor2}
	\end{align}
	with $p_{X^n}=p^n_X$ and $p_{V^n|U^n}=p^n_{V|U}$.
\end{enumerate}
 Additionally, we introduce an input cost function $c:\mathcal{U}\rightarrow[0,\infty)$. Unless  otherwise specified,  $\mathcal{X}$, $\mathcal{Y}$, $\mathcal{U}$, and $\mathcal{V}$ are assumed to have finite cardinalities.


\begin{definition}[Channel-Aware Optimal Transport With Common Randomness]\label{def:OTCw}
	With common randomness, we say distortion level $D$ is achievable through channel $p_{V|U}$ subject to input cost constraint $\Gamma$ if for any $\epsilon>0$ and all sufficiently large $n$, there exist seed distribution $p_Q$, encoding distribution $p_{U^n|X^nQ}$, and decoding distribution $p_{Y^n|V^nQ}$ such that 
	\begin{align}
		&p_{Y^n}=p^n_Y,\label{eq:constraint1}\\
		&\frac{1}{n}\sum\limits_{t=1}^n\mathbb{E}[c(U_t)]\leq \Gamma+\epsilon,\label{eq:constraint2}\\
		&\frac{1}{n}\sum\limits_{t=1}^n\mathbb{E}[d(X_t,Y_t)]\leq D+\epsilon.\label{eq:constraint3}		
	\end{align}
	The minimum achievable $D$ through channel $p_{V|U}$ subject to input cost constraint $\Gamma$ is denoted by $\underline{D}_J(\Gamma)$.
\end{definition}

\begin{definition}[Channel-Aware Optimal Transport Without Common Randomness]\label{def:OTCw/o}
	Without common randomness, we say distortion level $D$ is achievable through channel $p_{V|U}$ subject to input cost constraint $\Gamma$ if for any $\epsilon>0$ and all sufficiently large $n$, there exist  encoding distribution $p_{U^n|X^n}$ and decoding distribution $p_{Y^n|V^n}$ such that 
	\begin{align}
		&p_{Y^n}=p^n_Y,\\
		&\frac{1}{n}\sum\limits_{t=1}^n\mathbb{E}[c(U_t)]\leq \Gamma+\epsilon,\\
		&\frac{1}{n}\sum\limits_{t=1}^n\mathbb{E}[d(X_t,Y_t)]\leq D+\epsilon.
	\end{align}
	The minimum achievable $D$ through channel $p_{V|U}$ subject to input cost constraint $\Gamma$ without common randomness is denoted by $\overline{D}_J(\Gamma)$.
\end{definition}

The problems formulated in Definitions \ref{def:OTCw} and \ref{def:OTCw/o} are joint source-channel coding in nature.  This raises the question of whether the source-channel separation architecture---which first uses error correction codes to convert the channel into a rate-limited bit pipeline and then sends source codes through it---is optimal for them. To address this, we need to introduce the source-coding counterparts of these two problems. In source coding, the encoder outputs a message $M$ from alphabet $\mathcal{M}$, which is delivered to the decoder through an error-free bit pipeline subject to a rate constraint. Once again, we distinguish between two cases based on the availability of common randomness.
\begin{enumerate}
	\item With common randomness:  In this case, the encoder and decoder share a random seed $Q$, and their operations are specified by  conditional distributions $p_{M|X^nQ}$ and $p_{Y^n|MQ}$, respectively. The entire system is characterized by joint distribution $p_{X^nQMY^n}$, which factors as
	\begin{align}
		p_{X^nQMY^n}=p_{X^n}p_Qp_{M|X^nQ}p_{Y^n|MQ}\label{eq:factor3}
	\end{align}
	with $p_{X^n}=p^n_X$.
	
	\item Without common randomness: In this case, the encoder and decoder do not share a random seed, and their operations are specified by  conditional distributions $p_{M|X^n}$ and $p_{Y^n|M}$, respectively. The entire system is characterized by joint distribution $p_{X^nMY^n}$, which factors as
	\begin{align}
		p_{X^nMY^n}=p_{X^n}p_{M|X^n}p_{Y^n|M}\label{eq:factor4}
	\end{align}
	with $p_{X^n}=p^n_X$.
\end{enumerate}

\begin{definition}[Rate-Limited Optimal Transport With Common Randomness]
	With common randomness, we say distortion level $D$ is achievable subject to rate constraint $R$ if for any $\epsilon>0$ and all sufficiently large $n$, there exist seed distribution $p_Q$, encoding distribution $p_{M|X^nQ}$, and decoding distribution $p_{Y^n|MQ}$ such that  
	\begin{align}
		&p_{Y^n}=p^n_Y,\\
		&\frac{1}{n}\log|\mathcal{M}|\leq R+\epsilon,\\
		&\frac{1}{n}\sum\limits_{t=1}^n\mathbb{E}[d(X_t,Y_t)]\leq D+\epsilon.
	\end{align}
	The minimum achievable $D$ subject to rate constraint $R$ with common randomness is denoted by $\underline{D}'(R)$.
\end{definition}

\begin{definition}[Rate-Limited Optimal Transport Without Common Randomness]
	Without common randomness, we say distortion level $D$ is achievable subject to rate constraint $R$ if for any $\epsilon>0$ and all sufficiently large $n$, there exist  encoding distribution $p_{M|X^n}$ and decoding distribution $p_{Y^n|M}$ such that 
	\begin{align}
		&p_{Y^n}=p^n_Y,\\
		&\frac{1}{n}\log|\mathcal{M}|\leq R+\epsilon,\\
		&\frac{1}{n}\sum\limits_{t=1}^n\mathbb{E}[d(X_t,Y_t)]\leq D+\epsilon.
	\end{align}
	The minimum achievable $D$ subject to rate constraint $R$ without common randomness is denoted by $\overline{D}'(R)$.
\end{definition}

According to \cite[Sections III.A and III.B]{SLY15J2}, we have
\begin{align}
	\underline{D}'(R)=&\min\limits_{p_{XY}\in\Pi(p_X,p_Y)}\mathbb{E}[d(X,Y)]\\
	&\mbox{s.t.}\quad I(X;Y)\leq R,
\end{align}
and
\begin{align}
	\overline{D}'(R)=&\min\limits_{p_{XUY}\in\Omega(p_X,p_Y)}\mathbb{E}[d(X,Y)]\\
	&\mbox{s.t.}\quad \max\{I(X;W);I(Y;W)\}\leq R,
\end{align}
where $\Omega(p_X,p_Y)$ denotes the set of joint distributions $p_{XWY}$ with $p_{XY}\in\Pi(p_X,p_Y)$ and $X\leftrightarrow W\leftrightarrow Y$ forming a Markov chain.
Let 
\begin{align}
	&\underline{D}_S(\Gamma):=\underline{D}'(C(p_{V|U},\Gamma)),\\
	&\overline{D}_S(\Gamma):=\overline{D}'(C(p_{V|U},\Gamma)),
\end{align}
where $C(p_{V|U},\Gamma)$ is the capacity of channel $p_{V|U}$ subject to input cost constraint $\Gamma$, i.e.,
\begin{align}
	C(p_{V|U},\Gamma):=&\max\limits_{p_U}I(U;V)\\
	&\mbox{s.t.}\quad \mathbb{E}[c(U)]\leq\Gamma.
\end{align}
Clearly, $\underline{D}_S(\Gamma)$ and $\overline{D}_S(\Gamma)$ are respectively the minimum achievable distortions for separation-based schemes with and without common randomness. Note that the restriction of $\mathcal{X}$, $\mathcal{Y}$, $\mathcal{U}$, and $\mathcal{V}$ to finite cardinalities is not necessary for this statement. In particular, it remains valid for the  case where $p_X$ and $p_Y$ are Gaussian distributions and $p_{V|U}$ is an additive white Gaussian noise channel subject to an average input power constraint (which will be referred to simply as the Gaussian case  in the remainder of this paper).


\section{Main Results}\label{sec:mainresults}

Our first result indicates that the source-channel separation architecture is optimal when the encoder and decoder have access to unlimited common randomness. 
\begin{theorem}\label{thm:joint_with_common_randomness}
	We have
	\begin{align}
	\underline{D}_J(\Gamma)=\underline{D}_S(\Gamma).
	\end{align}
Henceforth, they will be simply written as $\underline{D}(\Gamma)$.
\end{theorem}
\begin{remark}
	Theorem \ref{thm:joint_with_common_randomness} holds for the Gaussian case as well.
\end{remark}
\begin{IEEEproof}
	It suffices to show $\underline{D}_J(\Gamma)\geq\underline{D}_S(\Gamma)$ since the other direction is trivially true. The details can be found in Appendix \ref{app:Theorem1}. 	
\end{IEEEproof}

As shown by the following two toy examples, the source-channel separation architecture may not be optimal when no common randomness is available.

\begin{enumerate}
	\item Binary case: Let $p_X=p_Y=\mathcal{B}(\frac{1}{2})$,  $p_{V|U}=\mathrm{BSC}(\theta)$\footnote{The binary symmetric channel with crossover probability $\theta$ is written as $\mathrm{BSC}(\theta)$. Its input-output relationship is given by $V=U\oplus N$, where $\oplus$ denotes  modulo-$2$ addition, and the channel noise $N\sim\mathcal{B}(\theta)$ is independent of the channel input $U$.} with $\theta\in[0,1]$, and $d(\cdot,\cdot)$ be the Hamming distortion measure $d_H(\cdot,\cdot)$ with $d_H(x,y):=0$ if $x=y$ and $d_H(x,y):=1$ otherwise. No input cost constraint is imposed, so 
	$\Gamma$ is omitted from the relevant expressions.
	
	 According to \cite[Example 1]{CYWSGT22},
	 \begin{align}
	 	&\underline{D}'(R)=H^{-1}_b((1-R)_+),\\
	 	&\overline{D}'(R)=2(1-H^{-1}_b((1-R)_+))H^{-1}_b((1-R)_+),
	 \end{align}
 where $H^{-1}_b(\cdot)$ denotes the inverse of the binary entropy function $H_b(\cdot)$ with its domain restricted to $[0,\frac{1}{2}]$. Since 
 \begin{align}
 	C(\mathrm{BSC}(\theta))=1-H_b(\theta),
 \end{align}
it follows that
\begin{align}
	&\underline{D}=\underline{D}'(C(\mathrm{BSC}(\theta)))=\theta,\\
	&\overline{D}_S=\overline{D}'(C(\mathrm{BSC}(\theta)))=2(1-\theta)\theta.\label{eq:binary_separation}
\end{align}

Consider an uncoded scheme with $U(t):=X(t)$ and $Y(t):=V(t)$ for $t=1,\ldots,n$. It is straightforward to verify that $Y(1), \ldots, Y(n)$ are i.i.d. with marginal distribution $\mathcal{B}(\frac{1}{2})$ and the  end-to-end distortion coincides with $\underline{D}$. As this uncoded scheme does not require common randomness yet achieves the minimum distortion even when unlimited common randomness is available, we must have
\begin{align}
	\overline{D}_J=\theta.\label{eq:binary_joint}
\end{align}
Comparing \eqref{eq:binary_separation} and \eqref{eq:binary_joint} shows 	
 $\overline{D}_J<\overline{D}_S$ for $\theta\in(0,\frac{1}{2})$.
	
	\item Gaussian case: Let $p_X=\mathcal{N}(\mu_X,\sigma^2_X)$, $p_Y=\mathcal{N}(\mu_Y,\sigma^2_Y)$, $p_{V|U}=\mathrm{AWGN}(1)$\footnote{The additive white Gaussian noise channel with unit noise variance is written as $\mathrm{AWGN}(1)$. Its input-output relationship is given by $V=U+N$, where the channel noise $N\sim\mathcal{N}(0,1)$ is independent of the channel input $U$.} with the input cost function given by $c(u)=u^2$, and $d(\cdot,\cdot)$ be the squared error distortion measure (i.e., $d(x,y)=\|x-y\|^2$).

	 According to \cite[Theorem 2]{XLCZ24},
	 \begin{align}
	 	&\underline{D}'(R)=(\mu_X-\mu_Y)^2+\sigma^2_X+\sigma^2_Y-2\sqrt{1-2^{-2R}}\sigma_X\sigma_Y,\\
	 	&\overline{D}'(R)=(\mu_X-\mu_Y)^2+\sigma^2_X+\sigma^2_Y-2(1-2^{-2R})\sigma_X\sigma_Y.
	 \end{align}
	Since 
	\begin{align}
		C(\mathrm{AWGN}(1),\Gamma)=\frac{1}{2}\log(\Gamma+1),
	\end{align}
it follows that 
	\begin{align}
		&\underline{D}(\Gamma)=\underline{D}'(C(\mathrm{AWGN}(1),\Gamma))=(\mu_X-\mu_Y)^2+\sigma^2_X+\sigma^2_Y-2\sqrt{\frac{\Gamma}{\Gamma+1}}\sigma_X\sigma_Y,\\
		&\overline{D}_S(\Gamma)=\overline{D}'(C(\mathrm{AWGN}(1),\Gamma))=(\mu_X-\mu_Y)^2+\sigma^2_X+\sigma^2_Y-\frac{2\Gamma}{\Gamma+1}\sigma_X\sigma_Y.\label{eq:Gaussian_separate}
	\end{align}
Consider an uncoded scheme with $U(t):=\frac{\sqrt{P}}{\sigma_X}(X(t)-\mu_X)$ and $Y(t):=\frac{\sigma_Y}{\sqrt{\Gamma+1}}V(t)+\mu_Y$ for $t=1,\ldots,n$. It is straightforward to verify that $Y(1), \ldots, Y(n)$ are i.i.d. with marginal distribution $\mathcal{N}(\mu_Y,\sigma^2_Y)$ and the end-to-end distortion coincides with $\underline{D}(\Gamma)$. As this uncoded scheme does not require common randomness yet achieves the minimum distortion even when unlimited common randomness is available, we must have
	\begin{align}
		\overline{D}_J(\Gamma)=(\mu_X-\mu_Y)^2+\sigma^2_X+\sigma^2_Y-2\sqrt{\frac{\Gamma}{\Gamma+1}}\sigma_X\sigma_Y.\label{eq:Gaussian_joint}
	\end{align}
	Comparing \eqref{eq:Gaussian_separate} and \eqref{eq:Gaussian_joint} shows  $\overline{D}_J(\Gamma)<\overline{D}_S(\Gamma)$ for $\Gamma>0$.
\end{enumerate}

The uncoded schemes in the above two toy examples are one-shot optimal in the sense that they achieve the asymptotic minimum distortion even when $n=1$. While these schemes are well-known in the classical source-channel communication setting, their implications differ significantly. 
In the classical setting, the source-channel separation architecture is optimal. The purpose of uncoded schemes in this context\footnote{See \cite[p. 147, Problems 2 and 3]{CK81} and \cite[Theorem 6]{GRV03} for the general source-channel matching conditions regarding the optimality of uncoded schemes.} is merely to demonstrate that no sophisticated coding is required to achieve the asymptotic information-theoretic limit for problems like transmitting a binary uniform source over a binary symmetric channel (see, e.g.,  \cite[p. 117]{Jelinek68} and \cite[Sec. 11.8]{McEliece77}) or a Gaussian source over an additive white Gaussian noise channel\footnote{In the classical setting, the uncoded scheme for transmitting a Gaussian source over an additive white Gaussian noise channel employs a linear MMSE decoder $Y(t):=\frac{\sqrt{\Gamma}\sigma_X}{\Gamma+1}V(t)+\mu_X$, which is generally different from the linear decoder $Y(t):=\frac{\sigma_Y}{\sqrt{\Gamma+1}}V(t)+\mu_Y$ designed to produce the desired reconstruction distribution $\mathcal{N}(\mu_Y,\sigma^2_X)$. } \cite{Goblick65}. 
In contrast, for the problem of channel-aware optimal transport  addressed in this paper, uncoded schemes highlight a different point: joint source-channel coding can strictly outperform separation-based schemes when the encoder and decoder do not have access to common randomness.



These two toy examples also provide insights into why the source-channel separation architecture is generally suboptimal when no common randomness is available. With the separation-based architecture, the decoder essentially receives a lossy coded representation of $X^n$. To produce the desired $Y^n$
from this representation, the decoder must perform a generative operation, which inevitably introduces additional distortion. 
In these two toy examples, the required generative operations happen to align naturally with the respective communication channels. This alignment makes it advantageous to integrate the two, thereby avoiding the double distortion penalty that would arise from separating them. 
In contrast, when common randomness is available,  the decoder can delegate virtually all of its functionality to the encoder while focusing solely on conditional entropy decoding based on the shared random seed \cite[Theorem 3]{LZCK22}. Since the decoder no longer performs any generative operations, there is no need to exploit the channel's stochasticity. As a result, the separation-based architecture becomes optimal in this case.

In general, the generative operations performed by the decoder may not fully align with the characteristics of the communication channel. As a result, uncoded schemes do not always deliver competitive performance. To address this, we propose a hybrid coding scheme that integrates uncoded and separation-based approaches, allowing them to effectively complement each other. It is worth mentioning that this hybrid coding scheme is not a straightforward extension of its counterpart developed for the classical source-channel communication problem 
\cite{MLK15} due to the extra distributional constraint on the reconstruction.

\begin{theorem}\label{thm:hybrid}
	We have
	\begin{align}
		\overline{D}_J(\Gamma)&\leq\mathbb{E}[d(X,Y)]
	\end{align}
for any $p_{ZU|X}$ and $p_{Y|ZV}$ such that the induced joint distribution $p_{XZUVY}:=p_Xp_{ZU|X}p_{V|U}p_{Y|ZV}$ is compatible with the given marginal $p_Y$ and satisfies
\begin{align}
&\mathbb{E}[c(U)]\leq\Gamma,\label{eq:condition1}\\
&\max\{I(X;Z), I(Y;Z)\}\leq I(Z;V).\label{eq:condition2}
\end{align}
Moreover, for the auxiliary random variable $Z$, there is no loss of optimality in imposing the cardinality bound  $|\mathcal{Z}|\leq |\mathcal{X}|+|\mathcal{Y}|+|\mathcal{V}|+2$ on its alphabet $\mathcal{Z}$.
\end{theorem}
\begin{remark}
	It can be shown via the standard discretization procedure that Theorem \ref{thm:hybrid} extends to the Gaussian case by restricting $p_{ZU|X}$ and $p_{Y|ZV}$ to be conditional Gaussian distributions.
\end{remark}

\begin{remark}
	Uncoded schemes correspond to setting $Z$ as a constant while separation-based schemes correspond to setting $Z:=(W,U)$ with $U$ independent of $(X,W)$.	
\end{remark}

\begin{IEEEproof}
	The details can be found in Appendix \ref{app:hybrid}.
\end{IEEEproof}

\section{Binary Case}\label{sec:binarycase}

In this section, we focus on the case where $p_X=p_Y=\mathcal{B}(\rho)$ with $\rho\in[0,\frac{1}{2}]$, $p_{V|U}=\mathrm{BSC}(\theta)$ with $\theta\in[0,\frac{1}{2}]$, and  $d(x,y)=d_H(x,y)$. 
No input cost constraint is imposed, so $\Gamma$ is omitted from the relevant expressions. Since the problem  is trivial when $\rho=0$ and degenerates to the first toy example in Section \ref{sec:mainresults} when  $\rho=\frac{1}{2}$,  we assume $\rho\in(0,\frac{1}{2})$ in the following discussion. Note that enforcing $p_X=p_Y$ corresponds to the perfect realism setting in rate-distortion-perception coding.

First consider the scenario where the encoder and decoder have access to unlimited common randomness.  According to \cite[Example 1]{CYWSGT22},
\begin{align}
	\underline{D}'(R)=\begin{cases}
		\hat{D}(R),&R\in[0,H_b(\rho)),\\
		0,&R\in[H_b(\rho),\infty),
	\end{cases}
\end{align}
where $\hat{D}(R)$ is the unique number in $(0,2(1-\rho)\rho]$ satisfying
\begin{align}
	2H_b(\rho)+\frac{2-2\rho-\hat{D}(R)}{2}\log\left(\frac{2-2\rho-\hat{D}(R)}{2}\right)+\hat{D}(R)\log\left(\frac{\hat{D}(R)}{2}\right)+\frac{2\rho-\hat{D}(R)}{2}\log\left(\frac{2\rho-\hat{D}(R)}{2}\right)=R.
\end{align}
Therefore, we have
\begin{align}
	\underline{D}=\underline{D}'(C(\mathrm{BSC}(\theta)))=\begin{cases}
		\hat{D}(1-H_b(\theta)),&\theta\in(H^{-1}_b(1-H_b(\rho)),\frac{1}{2}],\\
		0,&\theta\in[0,H^{-1}_b(1-H_b(\rho))].
	\end{cases}
\end{align}

Next consider the scenario where no common randomness is available to the encoder and decoder. We shall analyze the following schemes separately.

\begin{enumerate}
		\item Separation-based scheme: According to \cite[Example 1]{CYWSGT22},
		\begin{align}
			\overline{D}'(R)=2(1-\delta(R))\delta(R),
		\end{align}
	where
	\begin{align}
		\delta(R):=	H^{-1}_b((H_b(\rho)-R)_+).
	\end{align}
	Therefore, we have
	\begin{align}
		\overline{D}_S=\overline{D}'(C(\mathrm{BSC}(\theta)))=2(1-\delta(1-H_b(\theta)))\delta(1-H_b(\theta)).
	\end{align}
This separation-based scheme corresponds to setting $X:=W\oplus E$, $Y:=W\oplus\tilde{E}$, and $Z:=(W,U)$ with $W\sim\mathcal{B}(\frac{\rho-\delta(1-H_b(\theta))}{1-2\delta(1-H_b(\theta))})$, $E\sim\mathcal{B}(\delta(1-H_b(\theta)))$, $\tilde{E}\sim\mathcal{B}(\delta(1-H_b(\theta)))$,   and $U\sim\mathcal{B}(\frac{1}{2})$  in Theorem \ref{thm:hybrid} such that $W$, $E$, $\tilde{E}$, and $U$ are mutually independent and also independent of the channel noise $N$.
Clearly, 
\begin{align}
	I(X;Z)&=I(X;W)\nonumber\\
	&=H_b(\rho)-H_b(\delta(1-H_b(\theta)))\nonumber\\
	&=H_b(\rho)\wedge(1-H_b(\theta)).
\end{align}
Similarly, we also have
\begin{align}
	I(Y;Z)=H_b(\rho)\wedge(1-H_b(\theta)).
\end{align}
In view of the fact that
\begin{align}
	I(Z;V)=I(U;V)=1-H_b(\theta),
\end{align}
condition \eqref{eq:condition2} is  fulfilled. Finally, the end-to-end distortion indeed coincides with $\overline{D}_S$ because
\begin{align}
	\mathbb{E}[d_H(X,Y)]&=\mathbb{E}[E\oplus\tilde{E}]\nonumber\\
	&=2(1-\delta(1-H_b(\theta)))\delta(1-H_b(\theta)).
\end{align}

	\item Uncoded scheme: This scheme corresponds to setting $Z:=0$ and $U:=X$ in Theorem \ref{thm:hybrid}. Only $p_{Y|V}$ remains to be specified. It can be verified that
	\begin{align*}
		p_Y(1)=(1-(\rho*\theta))a+(\rho*\theta)(1-b)
	\end{align*}
and
\begin{align*}
	\mathbb{E}[d_H(X,Y)]=(1-\rho)((1-\theta)a+\theta(1-b))+\rho(\theta(1-a)+(1-\theta)b),
\end{align*}
where $a:=p_{Y|V}(1|0)$ and $b:=p_{Y|V}(0|1)$. Optimizing $\mathbb{E}[d_H(X,Y)]$ over $a\in[0,1]$ and $b\in[0,1]$ subject to the constraint $p_Y=\mathcal{B}(\rho)$ yields the minimum achievable distortion of the uncoded scheme:
\begin{align}
	\overline{D}_U:=&\min\limits_{a\in[0,1],b\in[0,1]}(1-\rho)((1-\theta)a+\theta(1-b))+\rho(\theta(1-a)+(1-\theta)b)\label{eq:objective}\\
	&\mbox{s.t.}\quad (1-(\rho*\theta))a+(\rho*\theta)(1-b)=\rho.\label{eq:constraint}
\end{align}  
It follows by the constraint in \eqref{eq:constraint} that
\begin{align}
	b=\frac{((1-\rho)(1-\theta)+\rho\theta)a+(1-2\rho)\theta}{\rho*\theta}.\label{eq:sub}
\end{align} 
Substituting \eqref{eq:sub} into the objective function in \eqref{eq:objective} gives
\begin{align}
(1-\rho)((1-\theta)a+\theta(1-b))+\rho(\theta(1-a)+(1-\theta)b)=\frac{2(1-\rho)\rho(1-2\theta)}{\rho*\theta}a+\frac{2(1-\rho)\rho\theta}{\rho*\theta}.
\end{align}
Since $\frac{2(1-\rho)\rho(1-2\theta)}{\rho*\theta}\geq 0$, the minimum value is attained at $a=0$, for which the corresponding  $b=\frac{(1-2\rho)\theta}{\rho*\theta}\in[0,1]$. Therefore, we have	
\begin{align}
	\overline{D}_U=\frac{2(1-\rho)\rho\theta}{\rho*\theta}.
\end{align}

Now we proceed to compare the uncoded scheme with the separation-based scheme. Note that $\overline{D}_S=0$ for $\theta\in[0,H^{-1}_b(1-H_b(\rho))]$ while $\left.\overline{D}_U\right|_{\theta=0}=0$ and $\overline{D}_U>0$ for $\theta\in(0,\frac{1}{2}]$. Thus, the separation-based scheme outperforms the uncoded scheme when $\theta$ is close to, but not equal to, zero. On the other hand, we have
\begin{align}
	&\left.\frac{\partial\overline{D}_S}{\partial\theta}\right|_{\theta=\frac{1}{2}}=0,\label{eq:symmetry}\\
	&\left.\frac{\partial\overline{D}_U}{\partial\theta}\right|_{\theta=\frac{1}{2}}=8(1-\rho)\rho^2>0,\label{eq:positiveslope}
\end{align}
where \eqref{eq:symmetry} can be inferred from the symmetry of $\overline{D}_S$ with respect to $\theta$ at $\theta=\frac{1}{2}$. Combing \eqref{eq:symmetry}, \eqref{eq:positiveslope}, and the fact $\left.\overline{D}_S\right|_{\theta=\frac{1}{2}}=\left.\overline{D}_U\right|_{\theta=\frac{1}{2}}=2(1-\rho)\rho$ shows that the uncoded scheme is more competitive than the separation-based scheme when when $\theta$ is close to, but not equal to, $\frac{1}{2}$.

	\item Hybrid coding scheme: Significant flexibility exists in the construction of such a scheme. We focus on a version that corresponds to setting $X:=W\oplus E_1\oplus E_2$, $Z:=(W,U_d)$, $U:=E_1\oplus U_d$, and $Y:=W\oplus\tilde{E}$ in Theorem \ref{thm:hybrid}. Here $W\sim\mathcal{B}(\tau)$, $E_1\sim\mathcal{B}(\delta_1)$, $E_2\sim\mathcal{B}(\delta_2)$, and $U_d\sim\mathcal{B}(\frac{1}{2})$ are mutually independent and also independent of the channel noise $N$; moreover, $\tilde{E}$ is conditionally independent of $(W,E_1,E_2,N)$ given $(U_d,V)$ with $\tilde{E}\sim\mathcal{B}(\alpha)$ if $U_d\oplus V=0$ and $\tilde{E}\sim\mathcal{B}(\beta)$ if $U_d\oplus V=1$. Note that $p_X$ is preserved by enforcing
	\begin{align}
			\tau*(\delta_1*\delta_2)=\rho.\label{eq:tau}
	\end{align}
To ensure $Y\sim\mathcal{B}(\rho)$, we need $\tilde{E}\sim\mathcal{B}(\delta_1*\delta_2)$, which boils down to requiring
\begin{align}
	(1-(\delta_1*\theta))\alpha+(\delta_1*\theta)\beta=\delta_1*\delta_2\label{eq:alphabeta1}
\end{align}
since  $U_d\oplus V=E_1\oplus N$. It is easy to verify that
\begin{align}
	I(X;Z)=I(X;W)=H_b(\rho)-H_b(\delta_1*\delta_2).
\end{align}
Similarly, we also have
\begin{align}
	I(Y;Z)=H_b(\rho)-H_b(\delta_1*\delta_2).
\end{align}
In view of the fact that
\begin{align}
	I(Z;V)=I(U_d;V)=1-H_b(\delta_1*\theta),
\end{align}	
condition \eqref{eq:condition2} becomes
\begin{align}
	H_b(\rho)-H_b(\delta_1*\delta_2)\leq 1-H_b(\delta_1*\theta),
\end{align}
which is fulfilled by choosing
\begin{align}
	\delta_2=\begin{cases}
		\frac{H_b^{-1}(H_b(\rho)+H_b(\delta_1*\theta)-1)-\delta_1}{1-2\delta_1},&H_b(\rho)-H_b(\delta_1)> 1-H_b(\delta_1*\theta),\\
		0,& H_b(\rho)-H_b(\delta_1)\leq 1-H_b(\delta_1*\theta).
	\end{cases}\label{eq:delta_2}
\end{align}
In light of \eqref{eq:tau}, 
\begin{align}
	\tau=\frac{\rho-(\delta_1*\delta_2)}{1-2(\delta_1*\delta_2)}.
\end{align}
With this construction, both $\delta_2$ and $\tau$ are determined by $\delta_1$. 
We restrict $\delta_1$ to $[0,\rho]$ to ensure they are well-defined.

The end-to-end distortion for this hybrid coding scheme is given by	
\begin{align}
		\mathbb{E}[d_H(X,Y)]&=\mathbb{E}[E_1\oplus E_2\oplus\tilde{E}]\nonumber\\
		&=p_{E_1E_2\tilde{E}}(0,1,0)+p_{E_1E_2\tilde{E}}(0,0,1)+p_{E_1E_2\tilde{E}}(1,0,0)+p_{E_1E_2\tilde{E}}(1,1,1)\nonumber\\
		&=(1-\delta_1)\delta_2p_{\tilde{E}|E_1}(0|0)+(1-\delta_1)(1-\delta_2)p_{\tilde{E}|E_1}(1|0)+\delta_1(1-\delta_2)p_{\tilde{E}|E_1}(0|1)+\delta_1\delta_2p_{\tilde{E}|E_1}(1|1),
	\end{align}
where
	\begin{align}
		&p_{\tilde{E}|E_1}(0|0)=(1-\theta)(1-\alpha)+\theta(1-\beta),\label{eq:conditionalprob1}\\
		&p_{\tilde{E}|E_1}(1|0)=(1-\theta)\alpha+\theta\beta,\label{eq:conditionalprob2}\\
		&p_{\tilde{E}|E_1}(0|1)=(1-\theta)(1-\beta)+\theta(1-\alpha),\label{eq:conditionalprob3}\\
		&p_{\tilde{E}|E_1}(1|1)=(1-\theta)\beta+\theta\alpha.\label{eq:conditionalprob4}
	\end{align}
Via some algebraic manipulations, we obtain
\begin{align}
	\mathbb{E}[d_H(X,Y)]=(1-\delta_1-\theta)(1-2\delta_2)\alpha-(\delta_1-\theta)(1-2\delta_2)\beta+(\delta_1*\delta_2).\label{eq:algebra}
\end{align}
In view of \eqref{eq:alphabeta1} and \eqref{eq:algebra}, the optimal choice of $(\alpha,\beta)$ can be determined by solving the following optimization problem:
\begin{align}
	&\min\limits_{\alpha\in[0,1],\beta\in[0,1]}(1-\delta_1-\theta)(1-2\delta_2)\alpha-(\delta_1-\theta)(1-2\delta_2)\beta+(\delta_1*\delta_2)\label{eq:etedistortion}\\
	&\mbox{s.t.}\quad (1-(\delta_1*\theta))\alpha+(\delta_1*\theta)\beta=\delta_1*\delta_2.\label{eq:alphabeta}
\end{align}
It follows by the constraint in \eqref{eq:alphabeta} that
\begin{align}
	\beta=\frac{\delta_1*\delta_2-(1-(\delta_1*\theta))\alpha}{\delta_1*\theta}.\label{eq:beta}
\end{align}
Substituting \eqref{eq:beta} into the objective function in \eqref{eq:etedistortion}  yields
\begin{align}
	&(1-\delta_1-\theta)(1-2\delta_2)\alpha-(\delta_1-\theta)(1-2\delta_2)\beta+(\delta_1*\delta_2)\nonumber\\
	&=\frac{2(1-\delta_1)\delta_1(1-2\delta_1)(1-2\theta)}{\delta_1*\theta}\alpha+\frac{2(\delta_1*\delta_2)((1-\delta_1-\delta_2)\theta+\delta_1\delta_2)}{\delta_1*\theta}.
\end{align}
Since $\frac{2(1-\delta_1)\delta_1(1-2\delta_1)(1-2\theta)}{\delta_1*\theta}\geq 0$, the minimum value is attained at $\alpha=0$, for which the corresponding\footnote{Note that \eqref{eq:delta_2} implies $\delta_2\leq\theta$.} $\beta=\frac{\delta_1*\delta_2}{\delta_1*\theta}\in[0,1]$. The resulting end-to-end distortion is 
\begin{align}
	\mathbb{E}[d_H(X,Y)]=\overline{D}_H(\delta_1):=\frac{2(\delta_1*\delta_2)((1-\delta_1-\delta_2)\theta+\delta_1\delta_2)}{\delta_1*\theta}.
\end{align}
Optimizing $\overline{D}_H(\delta_1)$ over $\delta_1$ yields the minimum achievable distortion for this hybrid coding scheme\footnote{In this derivation, we have implicitly assumed that $\theta>0$, and consequently, $\delta_1*\theta>0$. It can be verified that when $\theta=0$, the minimum end-to-end distortion for this hybrid coding scheme is zero, achieved by setting $\delta_1=\delta_2=\alpha=0$ (with the choice of $\beta$ being irrelevant).}:
\begin{align}
	\overline{D}_H&:=\min\limits_{\delta_1\in[0,\rho]}\overline{D}_H(\delta_1)\nonumber\\
	&=\min\limits_{\delta_1\in[0,\rho]}\frac{2(\delta_1*\delta_2)((1-\delta_1-\delta_2)\theta+\delta_1\delta_2)}{\delta_1*\theta}.\label{eq:maximizerdelta}
\end{align}


This hybrid coding scheme simplifies to the separation-based scheme when $\delta_1=0$ and
to the uncoded scheme
when $\delta_1=\rho$. Consequently, it performs at least on par with both of these schemes. Next we shall show that the hybrid coding scheme can strictly outperform the uncoded scheme for $\theta\in[\frac{\rho^2}{2\rho^2-\rho+1},\frac{1}{2})$. Let
\begin{align}
	\delta'_1(\theta):=\begin{cases}
		\delta'_1, & H_b(\rho)> 1-H_b(\theta),\\
		0, &H_b(\rho)\leq 1-H_b(\theta),		
	\end{cases}
\end{align}
where $\delta'_1$ is the unique number in $(0,\rho)$ satisfying\footnote{Let $\psi(\delta):=H_b(\delta*\theta)-H_b(\delta)$. We have $\psi(0)=H_b(\theta)>1-H_b(\rho)$ and $\psi(\rho)=H_b(\rho*\theta)-H_b(\rho)<1-H_b(\rho)$; moreover,  $\frac{\mathrm{d}\psi(\delta)}{\mathrm{d}\delta}=(1-2\theta)\log(\frac{(1-(\delta*\theta))}{\delta*\theta})-\log(\frac{1-\delta}{\delta})$ and $\frac{\mathrm{d}^2\psi(\delta)}{\mathrm{d}\delta^2}=\frac{(1-\theta)\theta\log(e)}{(1-\delta)\delta(1-(\delta*\theta))(\delta*\theta)}$. Since $\left.\frac{\mathrm{d}\psi(\delta)}{\mathrm{d}\delta}\right|_{\delta=\frac{1}{2}}=0$ and $\frac{\mathrm{d}^2\psi(\delta)}{\mathrm{d}\delta^2}>0$ for $\delta\in(0,\frac{1}{2}]$, 
it follows that $\psi(\delta)$ is strictly decreasing in $\delta$ for $\delta\in[0,\frac{1}{2}]$. As a consequence, there must exist a unique number $\delta'_1\in(0,\rho)$ satisfying $\psi(\delta'_1)=1-H_b(\rho)$.} $H_b(\delta'_1*\theta)-H_b(\delta'_1)=1-H_b(\rho)$.
Consider a simplified, though potentially suboptimal, hybrid coding scheme with $\delta_1=\delta'_1(\theta)$ and $\delta_2=0$. The corresponding end-to-end distortion is given by
\begin{align}
	\overline{D}'_H:=\frac{2(1-\delta'_1(\theta))\delta'_1(\theta)\theta}{\delta'_1(\theta)*\theta}.
\end{align}
Let $\phi(\delta):=\frac{2(1-\delta)\delta\theta}{\delta*\theta}$. It can be verified that
\begin{align}
	\frac{\mathrm{d}\phi(\delta)}{\mathrm{d}\delta}=\frac{-\theta(1-2\theta)\delta^2-2\theta^2\delta+\theta^2}{(\delta*\theta)^2}.
\end{align}
The numerator $-\theta(1-2\theta)\delta^2-2\theta^2\delta+\theta^2$, as a concave quadratic function of $\delta$, has a negative root at $-\frac{\sqrt{\theta-\theta^2}+\theta}{1-2\theta}$ and a positive root at $\frac{\sqrt{\theta-\theta^2}-\theta}{1-2\theta}$.
For $\theta\in[\frac{\rho^2}{2\rho^2-\rho+1},\frac{1}{2})$, in view of the fact $2\rho^2-2\rho+1>0$, we have
\begin{align}
	-\theta(1-2\theta)\rho^2-2\theta^2\rho+\theta^2&=(-\rho^2+(2\rho^2-2\rho+1)\theta)\theta\nonumber\nonumber\\
	&\geq \left(-\rho^2+(2\rho^2-2\rho+1)\frac{\rho^2}{2\rho^2-\rho+1}\right)\theta\nonumber\\
	&=0,
\end{align}
which implies $\rho\leq\frac{\sqrt{\theta-\theta^2}-\theta}{1-2\theta}$; it follows that
\begin{align}
	\frac{\mathrm{d}\phi(\delta)}{\mathrm{d}\delta}>0,\quad\delta\in(0,\rho),
\end{align}
and consequently $\phi(\delta'_1(\theta))<\phi(\rho)$, i.e., $\overline{D}'_H<\overline{D}_U$.
\end{enumerate}

\begin{figure}[htbp]
	\centerline{\includegraphics[width=12cm]{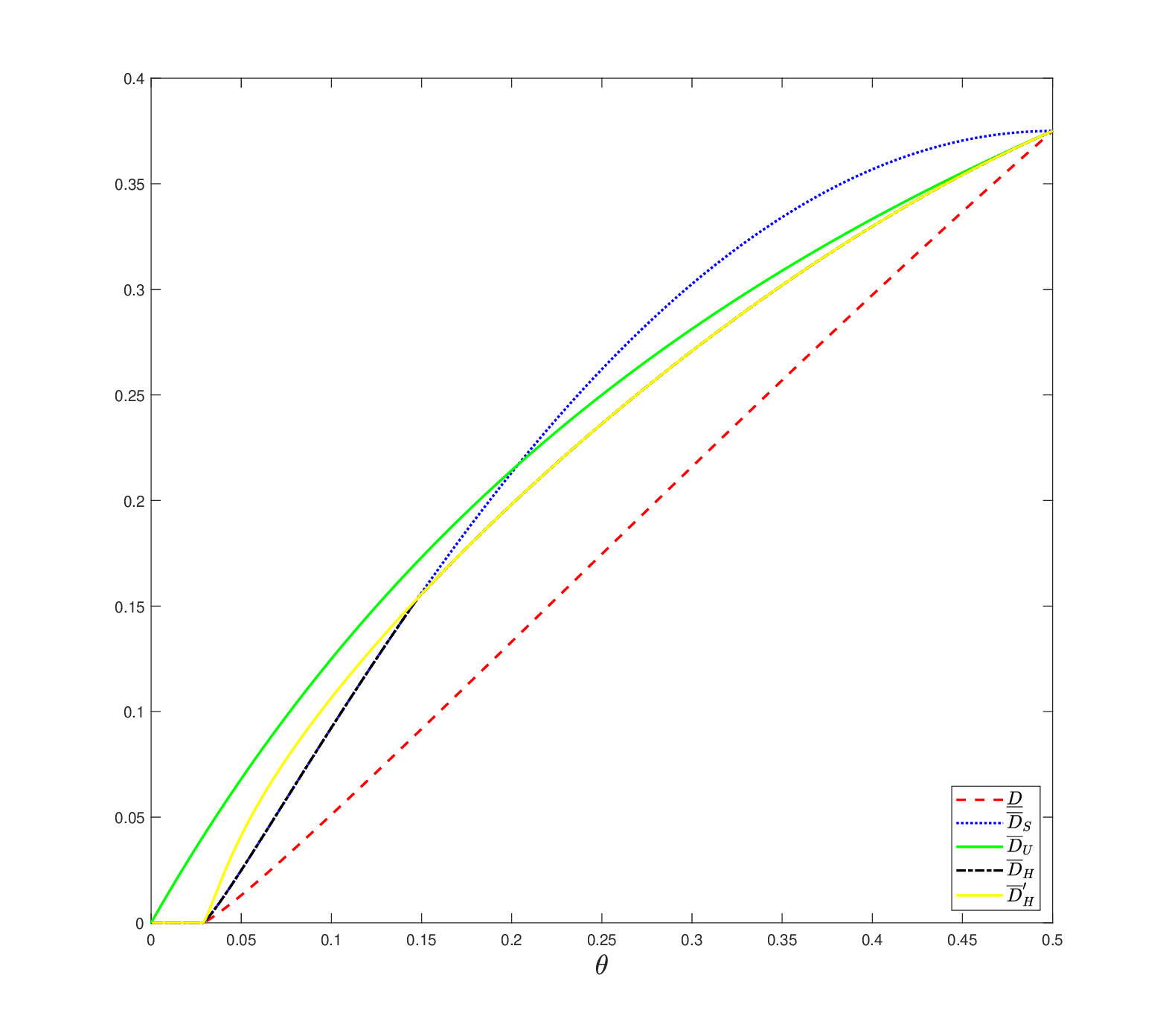}} \caption{Plots of $\underline{D}$, $\overline{D}_S$, $\overline{D}_U$,  $\overline{D}_H$, and $\overline{D}'_H$ against $\theta$ for the binary case with $\rho=\frac{1}{4}$.}
	\label{fig:binary} 
\end{figure}


\begin{figure}[htbp]
	\centerline{\includegraphics[width=12cm]{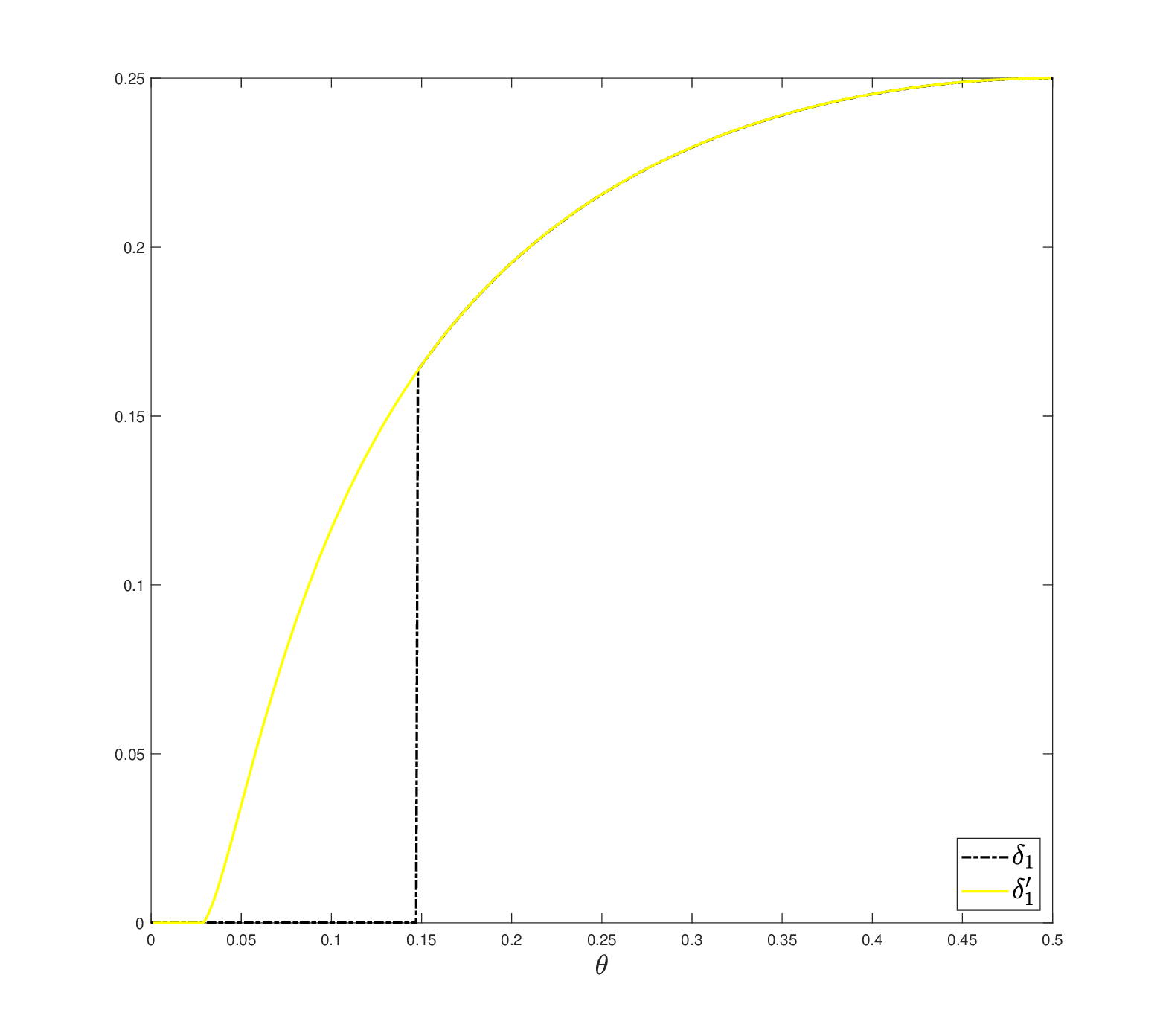}} \caption{Plots of $\delta_1(\theta)$ and $\delta'_1(\theta)$ for the binary case with $\rho=\frac{1}{4}$.}
	\label{fig:theta_delta1'} 
\end{figure}

\begin{figure}[htbp]
	\centerline{\includegraphics[width=12cm]{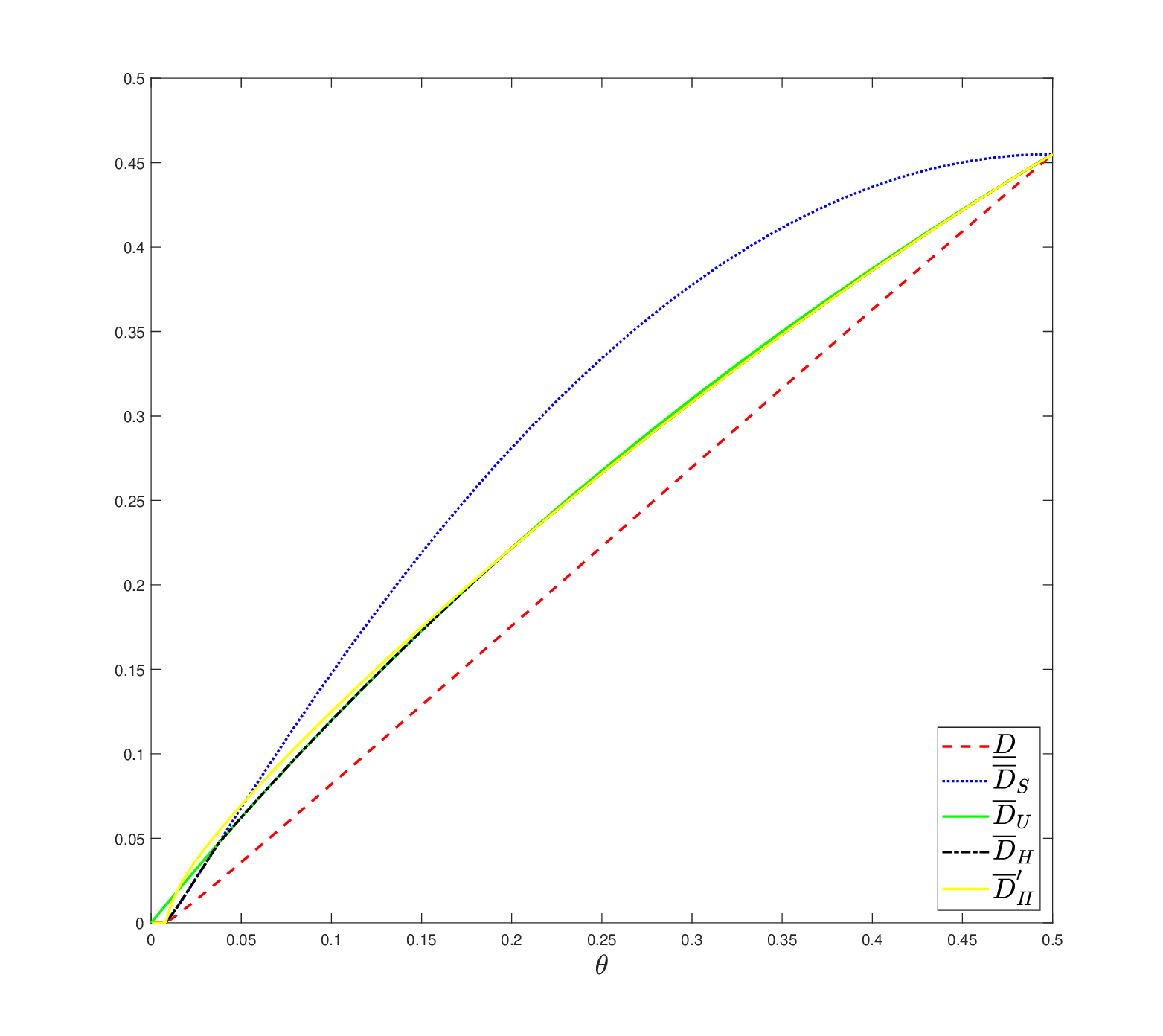}} \caption{Plots of $\underline{D}$, $\overline{D}_S$, $\overline{D}_U$,   $\overline{D}_H$, and $\overline{D}'_H$ against $\theta$ for the binary case with $\rho=\frac{7}{20}$.}
	\label{fig:D35} 
\end{figure}

\begin{figure}[htbp]
	\centerline{\includegraphics[width=12cm]{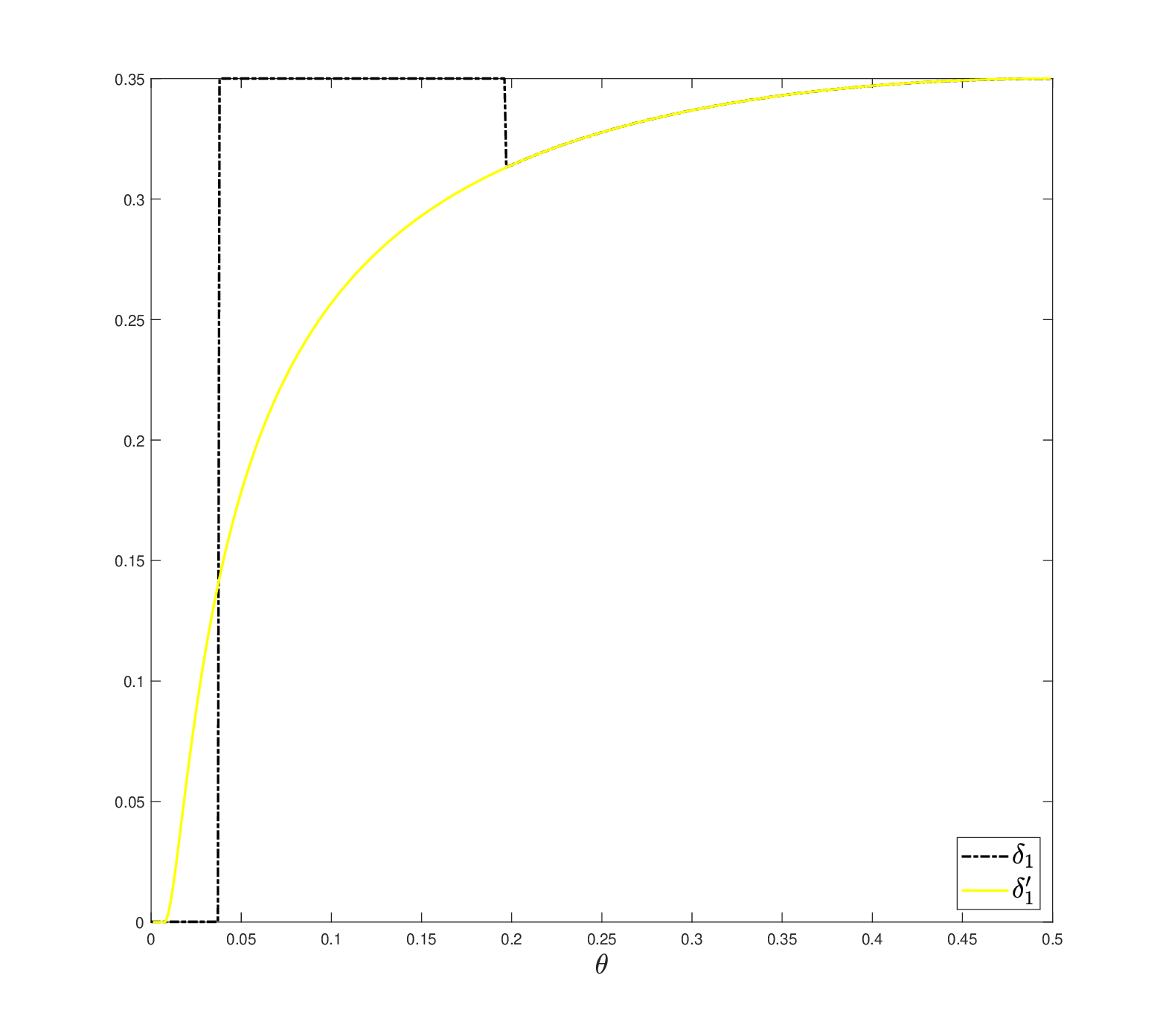}} \caption{Plots of $\delta_1(\theta)$ and $\delta'_1(\theta)$ for the binary case with $\rho=\frac{7}{20}$.}
	\label{fig:delta35} 
\end{figure}

In Fig. \ref{fig:binary}, we plot $\underline{D}$, $\overline{D}_S$, $\overline{D}_U$,  $\overline{D}_H$, and $\overline{D}'_H$ against $\theta$ for the special case of $\rho=\frac{1}{4}$. It can be seen that the separation-based scheme outperforms the uncoded scheme in the small-$\theta$ regime but becomes less competitive as $\theta$ increases. The hybrid coding scheme dominates both the uncoded scheme and separation-based schemes, yet it
still falls short of attaining $\underline{D}$---the minimum achievable distortion in the scenario with the availability of unlimited common randomness to the encoder and decoder. Another observation is that $\overline{D}_H$ appears to coincide with $\overline{D}_S$ when $\theta$ is below a certain threshold, $\hat{\theta} \approx 0.148$, and with $\overline{D}'_H$ when $\theta$ exceeds $\hat{\theta}$. To confirm this, we plot the minimizer of \eqref{eq:maximizerdelta}, denoted as $\delta_1(\theta)$, in Fig. \ref{fig:theta_delta1'}. The plot shows that $\delta_1(\theta)$ equals zero for $\theta \leq \hat{\theta}$ and matches $\delta'_1(\theta)$ for $\theta \geq \hat{\theta}$. This indicates that the optimized hybrid coding scheme operates in two distinct modes, the separation-based scheme and the simplified form, depending on the channel crossover probability.

In Fig. \ref{fig:D35}, we examine the special case of $\rho=\frac{7}{20}$. It can be seen that
there are two thresholds, $\hat{\theta}_1\approx 0.037$ and $\hat{\theta}_2\approx 0.197$.  Specifically,  $\overline{D}_H=\overline{D}_S$ for $\theta\in[0,\hat{\theta}_1]$, $\overline{D}_H=\overline{D}_U$ for $\theta\in[\hat{\theta}_1,\hat{\theta}_2]$, and $\overline{D}_H=\overline{D}'_H$ for $\theta\in[\hat{\theta}_2,\frac{1}{2}]$.
This behavior is further corroborated  in Fig. \ref{fig:delta35}, where $\delta_1(\theta)$ intially remains at zero, jumps to $\rho$ at $\theta=\hat{\theta}_1$, and stays constant until $\theta=\hat{\theta}_2$, after which it follows it follows the trajectory of $\delta'_1(\theta)$. Consequently, the optimized hybrid coding scheme operates in three distinct modes: it begins with the separation-based scheme in the in the small-$\theta$ regime, transition to the uncoded scheme in the intermediate-$\theta$ regime, and ultimately adopts the simplified form in the large-$\theta$ regime.

\section{Gaussian Case}\label{sec:Gaussiancase}

This section is devoted to the case where $X:=(X_1,\ldots,X_L)^T\sim\mathcal{N}(\mu_X,\Sigma_X)$, $Y:=(Y_1,\ldots,Y_L)^T\sim\mathcal{N}(\mu_Y,\Sigma_Y)$, $p_{V|U}=\mathrm{AWGN}(1)$ with the input cost function given by $c(u)=u^2$, and  $d(x,y)=\|x-y\|^2$. For simplicity, we focus on the perfect realism setting, namely, $\mu_X=\mu_Y=\mu$ and $\Sigma_X=\Sigma_Y=\Sigma$. Without loss of generality, it is assumed that $\mu=0$ since the mean can always be subtracted at the encoder and added back at the decoder. Moreover, it suffices to consider the scenario where $\Sigma=\mathrm{diag}(\lambda_1,\ldots,\lambda_L)$ with $\lambda_1\geq\ldots\geq\lambda_L>0$ since $\Sigma$ can be diagonalized using eigenvalue decomposition and the squared error measure remains invariant under unitary transformations. Note that the problem is covered by the second toy example in Section \ref{sec:mainresults} when $L=1$. Hence, we assume $L\geq 2$ in the following discussion.


First consider the scenario where the encoder and decoder have access to unlimited common randomness.  According to \cite[Corollary 1]{QSCKYSGT24},
\begin{align}
	\underline{D}'(R)=2\sum\limits_{\ell=1}^L(\lambda_{\ell}-\gamma_{\ell}(R)),
\end{align}
where
\begin{align}
	\gamma_{\ell}(R):=\begin{cases}
	\frac{-1+\sqrt{1+4\kappa^2(R)\lambda^2_{\ell}}}{2\kappa(R)},& R>0,\\
	0,& R=0,
	\end{cases}
	\quad\ell=1,\ldots,L,
\end{align}
with $\kappa(R)$ being the unique positive number satisfying
\begin{align}
	\frac{1}{2}\sum\limits_{\ell=1}^L\log\left(\frac{\lambda^2_{\ell}}{\lambda^2_{\ell}-\gamma^2_{\ell}(R)}\right)=R.
\end{align}
Therefore, we have
\begin{align}
	\underline{D}(\Gamma)=\underline{D}'(C(\mathrm{AWGN}(1),\Gamma))=2\sum\limits_{\ell=1}^L\left(\lambda_{\ell}-\gamma_{\ell}\left(\frac{1}{2}\log(\Gamma+1)\right)\right).
\end{align}

Next consider the scenario where no common randomness is available to the encoder and decoder. We shall analyze the following schemes separately.
\begin{enumerate}
	\item Separation-based scheme: According to \cite[Theorem 6]{CYWSGT22},
	\begin{align}
		\overline{D}'(R)=2D(R),
	\end{align}
where $D(R)$ is the distortion-rate function for the vector Gaussian source with covariance matrix $\mathrm{diag}(\lambda_1,\ldots,\lambda_L)$ under the squared distortion measure. By the reverse waterfilling formula \cite[Theorem 13.3.3]{CT91}, 
\begin{align}
	D(R)=2\sum\limits_{\ell=1}^L(\omega^{(S)}(R)\wedge\lambda_{\ell})
\end{align}
with $\omega^{(S)}(R)$ being the unique number in $(0,\lambda_1]$ satisfying
\begin{align}
	\frac{1}{2}\sum\limits_{\ell=1}^L\log\left(\frac{\lambda_{\ell}}{\omega^{(S)}(R)\wedge\lambda_{\ell}}\right)=R.\label{eq:omegaR}
\end{align}
Therefore, we have
	\begin{align*}
		\overline{D}_S(\Gamma)=\overline{D}'(C(\mathrm{AWGN}(1)),\Gamma)=		
		2\sum\limits_{\ell=1}^L\left(\omega^{(S)}\left(\frac{1}{2}\log(\Gamma+1)\right)\wedge\lambda_{\ell}\right).
	\end{align*}
This separation-based scheme corresponds to setting $X:=W+E$, $Y:=W+\tilde{E}$, and $Z:=(W,U)$ with $W:=(W_1,\ldots,W_L)^T\sim\mathcal{N}(0,\mathrm{diag}(\lambda_1-\delta^{(S)}_1(\Gamma),\ldots,\lambda_L-\delta^{(S)}_L(\Gamma)))$, $E:=(E_1,\ldots,E_L)^T\sim\mathcal{N}(0,\mathrm{diag}(\delta^{(S)}_1(\Gamma),\ldots,\delta^{(S)}_L(\Gamma)))$, $\tilde{E}:=(\tilde{E}_1,\ldots,\tilde{E}_L)^T\sim\mathcal{N}(0,\mathrm{diag}(\delta^{(S)}_1(\Gamma),\ldots,\delta^{(S)}_L(\Gamma)))$,   and $U\sim\mathcal{N}(0,\Gamma)$  in Theorem \ref{thm:hybrid} such that $W$, $E$, $\tilde{E}$,  and $U$ are mutually independent and also independent of the channel noise $N$, where
\begin{align}
	\delta^{(S)}_{\ell}(\Gamma):=\omega^{(S)}\left(\frac{1}{2}\log(\Gamma+1)\right)\wedge\lambda_{\ell},\quad\ell=1,\ldots,L.\label{eq:delta}
\end{align}
Clearly, condition \eqref{eq:condition1} is fulfilled since $\mathbb{E}[U^2]=\Gamma$. Moreover, it can be verified that
\begin{align}
	I(X;Z)&=I(X;W)\nonumber\\
	&=\frac{1}{2}\sum\limits_{\ell=1}^L\log\left(\frac{\lambda_{\ell}}{\delta^{(S)}_{\ell}(\Gamma)}\right)\nonumber\\
	&=\frac{1}{2}\log(\Gamma+1),\label{eq:total}
\end{align}
where the last equality is due to \eqref{eq:omegaR} and \eqref{eq:delta}. Similarly, we also have
\begin{align}
	I(Y;Z)=\frac{1}{2}\log(\Gamma+1).
\end{align}
In view of the fact that
\begin{align}
	I(Z;V)=I(U;V)=\frac{1}{2}\log(\Gamma+1),
\end{align}
condition \eqref{eq:condition2} is fulfilled as well. Finally, the end-to-end distortion indeed coincides with $\overline{D}_S(\Gamma)$ because
\begin{align}
	\mathbb{E}[\|X-Y\|^2]&=\mathbb{E}[\|E-\tilde{E}\|^2]\nonumber\\
	&=2\sum\limits_{\ell=1}^L\delta^{(S)}_{\ell}(\Gamma)\nonumber\\
	&=2\sum\limits_{\ell=1}^L\left(\omega^{(S)}\left(\frac{1}{2}\log(\Gamma+1)\right)\wedge\lambda_{\ell}\right).
\end{align}


	\item Uncoded scheme: We apply the uncoded scheme from the second toy example in Section \ref{sec:mainresults} to the component with the largest variance and regenerate the remaining components from scratch. This corresponds to setting $Z:=0$, $U:=\sqrt{\frac{\Gamma}{\lambda_1}}X_1$, $Y_1:=\sqrt{\frac{\lambda_1}{\Gamma+1}}V$, and letting $(Y_2,\ldots,Y_L)^T\sim\mathcal{N}(0,\mathrm{diag}(\lambda_2,\ldots,\lambda_L))$ be independent of $(X,U,V,Y_1)$ 	
	 in Theorem \ref{thm:hybrid}. The resulting end-to-end distortion is given by
	 \begin{align}
	 	\overline{D}_U(\Gamma)&:=\mathbb{E}[\|X-Y\|^2]\nonumber\\
	 	&=\mathbb{E}[(X_1-Y_1)^2]+2\sum\limits_{\ell=2}^L\lambda_{\ell}\nonumber\\
	 	&=\mathbb{E}\left[\left(X_1-\sqrt{\frac{\lambda_1}{\Gamma+1}}\left(\sqrt{\frac{\Gamma}{\lambda_1}}X_1+N\right)\right)^2\right]+2\sum\limits_{\ell=2}^L\lambda_{\ell}\nonumber\\
	 	&=-2\sqrt{\frac{\Gamma}{\Gamma+1}}\lambda_1+2\sum\limits_{\ell=1}^L\lambda_{\ell}.
	 \end{align}
 As shown in Appendix \ref{app:linear}, this uncoded scheme achieves the minimum distortion among all linear schemes.
	
	It is instructive to compare this uncoded scheme with the separation-based scheme. 
	Let $L_1:=\max\{\ell\in\{1,\ldots,L\}:\lambda_{\ell}=\lambda_1\}$. When\footnote{We set $\lambda_{L_1+1}:=0$ if $L_1=L$.}
	\begin{align}
		\frac{\lambda_1}{(\Gamma+1)^{\frac{1}{L_1}}}\geq\lambda_{L_1+1},\label{eq:lowerpower}
	\end{align}
	we have
	\begin{align}
		\omega^{(S)}\left(\frac{1}{2}\log(\Gamma+1)\right)=\frac{\lambda_1}{(\Gamma+1)^{\frac{1}{L_1}}}
	\end{align}
	  and consequently
	\begin{align}
		\overline{D}_S(\Gamma)=\frac{2L_1\lambda_1}{(1+\Gamma)^{\frac{1}{L_1}}}+2\sum\limits_{\ell=L_1+1}^L\lambda_{\ell}.
	\end{align}
	It can be verified that for $\Gamma$ sufficiently close to zero, 
	 condition \eqref{eq:lowerpower} is satisfied and 
	\begin{align}
		\overline{D}_S(\Gamma)-\overline{D}_U(\Gamma)&=\frac{2L_1\lambda_1}{(\Gamma+1)^{\frac{1}{L_1}}}+2\sqrt{\frac{\Gamma}{\Gamma+1}}\lambda_1-2L_1\lambda_1\nonumber\\
		&=2\sqrt{\Gamma}\lambda_1+o(\sqrt{\Gamma}).
	\end{align}
Thus, the uncoded scheme outperforms the separation-based scheme in the low-power regime.
One the other hand, the uncoded scheme is strictly inferior to the separation-based scheme in the high-power regime since
\begin{align}
	&\lim\limits_{\Gamma\rightarrow\infty}\overline{D}_S(\Gamma)=0,\\
	&\lim\limits_{\Gamma\rightarrow\infty}\overline{D}_U(\Gamma)=2\sum\limits_{\ell=2}^L\lambda_{\ell}>0.
\end{align}



	\item Hybrid coding scheme: We consider a hybrid coding scheme comprising an analog part and a digital part. The analog part employs the uncoded scheme for the component with the largest variance while the digital part utilizes the speration-based scheme for the remaining components. We allocate power 
	$(1-\alpha)\Gamma$ to the analog part and power $\alpha\Gamma$ to the digital part. This scheme corresponds to setting  	
	$(X_2,\ldots,X_L)^T:=(W_2,\ldots,W_L)^T+(E_2,\ldots,E_L)^T$, $(Y_2,\ldots,Y_L)^T:=(W_2,\ldots,W_L)^T+(\tilde{E}_2,\ldots,\tilde{E}_L)^T$, $Z:=(W_2,\ldots,W_L,U_d)$, $U:=\sqrt{\frac{(1-\alpha)\Gamma}{\lambda_1}}X_1+U_d$, and $Y_1:=\sqrt{\frac{\lambda_1}{(1-\alpha)\Gamma+1}}(V-U_d)$ with $(W_2,\ldots,W_L)^T\sim\mathcal{N}(0,\mathrm{diag}(\lambda_2-\delta^{(H)}_2(\Gamma,\alpha),\ldots,\lambda_L-\delta^{(H)}_L(\Gamma,\alpha)))$,  $(E_2,\ldots,E_L)^T\sim\mathcal{N}(0,\mathrm{diag}(\delta^{(H)}_2(\Gamma,\alpha),\ldots,\delta^{(H)}_L(\Gamma,\alpha)))$, $(\tilde{E}_2,\ldots,\tilde{E}_L)^T\sim\mathcal{N}(0,\mathrm{diag}(\delta^{(H)}_2(\Gamma,\alpha),\ldots,\delta^{(H)}_L(\Gamma,\alpha)))$, and $U_d\sim\mathcal{N}(0,\alpha\Gamma)$	
	in Theorem \ref{thm:hybrid} such that $(W_2,\ldots,W_L)^T$, $(E_2,\ldots,E_L)^T$, $(\tilde{E}_2,\ldots,\tilde{E}_L)^T$,  and $U_d$ are mutually independent and also independent of $(X_1,N)$, where
	\begin{align}
		\delta^{(H)}_{\ell}(\Gamma,\alpha):=\omega^{(H)}(\Gamma,\alpha)\wedge\lambda_{\ell},\quad \ell=2,\ldots,L,
		\end{align}
	with $\omega^{(H)}(\Gamma,\alpha)$ being the unique number in $(0,\lambda_2]$ satisfying
	\begin{align}
		\prod\limits_{\ell=2}^L\left(\frac{\lambda_{\ell}}{\omega^{(H)}(\Gamma,\alpha)\wedge\lambda_{\ell}}\right)=\frac{\Gamma+1}{(1-\alpha)\Gamma+1}.\label{eq:power}
	\end{align}
Condition \eqref{eq:condition1} is fulfilled since
\begin{align}
	\mathbb{E}[U^2]=\mathbb{E}\left[\left(\sqrt{\frac{(1-\alpha)\Gamma}{\lambda_1}}X_1+U_d\right)^2\right]=\Gamma.
\end{align} 
Moreover, it can be verified that
\begin{align}
	I(X;Z)&=I(X_2,\ldots,X_L;W_2,\ldots,W_L)\nonumber\\
	&=\frac{1}{2}\sum\limits_{\ell=2}^L\log\left(\frac{\lambda_{\ell}}{\delta^{(H)}_{\ell}(\Gamma,\alpha)}\right)\nonumber\\
	&=\frac{1}{2}\log\left(\frac{\Gamma+1}{(1-\alpha)\Gamma+1}\right),
\end{align}
where the last equality is due to \eqref{eq:power}. Similarly, we also have
\begin{align}
I(Y;Z)=\frac{1}{2}\log\left(\frac{\Gamma+1}{(1-\alpha)\Gamma+1}\right).
\end{align}
In view of the fact that
\begin{align}
	I(Z;V)=I(U_d;V)=\frac{1}{2}\log\left(\frac{\Gamma+1}{(1-\alpha)\Gamma+1}\right),
\end{align}
condition \eqref{eq:condition2} is fulfilled as well. The end-to-end distortion for this hybrid coding scheme is given by 
\begin{align}
	D_H(\Gamma,\alpha)&:=\mathbb{E}[\|X-Y\|^2]\\
	&=\mathbb{E}[(X_1-Y_1)^2]+\sum\limits_{\ell=2}^L\mathbb{E}[(X_{\ell}-Y_{\ell})^2]\nonumber\\
	&=\mathbb{E}\left[\left(X_1-\sqrt{\frac{\lambda_1}{(1-\alpha)\Gamma+1}}\left(\sqrt{\frac{(1-\alpha)\Gamma}{\lambda_1}}X_1+N\right)\right)^2\right]+\sum\limits_{\ell=2}^L\mathbb{E}[(E_{\ell}-\tilde{E}_{\ell})^2]\nonumber\\
	&=2\left(\left(1-\sqrt{\frac{(1-\alpha)\Gamma}{(1-\alpha)\Gamma+1}}\right)\lambda_1+\sum\limits_{\ell=2}^L\delta^{(H)}_{\ell}(\Gamma,\alpha)\right).
\end{align}
Optimizing $D_H(\Gamma,\alpha)$ over $\alpha$ yields the minimum achievable distortion for this hybrid coding scheme:
\begin{align}
	\overline{D}_H(\Gamma)&:=\min\limits_{\alpha\in[0,1]}\overline{D}_H(\Gamma,\alpha)\nonumber\\
	&=\min\limits_{\alpha\in[0,1]}2\left(\left(1-\sqrt{\frac{(1-\alpha)\Gamma}{(1-\alpha)\Gamma+1}}\right)\lambda_1+\sum\limits_{\ell=2}^L\delta^{(H)}_{\ell}(\Gamma,\alpha)\right).\label{eq:minimizer}
\end{align}

Note that this hybrid coding scheme degenerates to the uncoded scheme when $\alpha=0$. Moreover, with a suitable choice of $\alpha$, it outperforms the separation-based scheme as shown below. To avoid confusion,  the random variables associated the hybrid coding scheme are labeled using the supercript $(H)$ while  those associated with the separation-based scheme are labeled using the superscript $(S)$. We choose\footnote{This choice of $\alpha$ might not be optimal, but it is sufficient to demonstrate the superiority of the hybrid coding scheme over the separation-based scheme.} $\alpha$ such that
\begin{align}
	\frac{1}{2}\log\left(\frac{\Gamma+1}{(1-\alpha)\Gamma+1}\right)=\sum\limits_{\ell=2}^LI(X^{(S)}_{\ell};W^{(S)}_{\ell}),\label{eq:alpha}
\end{align}
which, in view of \eqref{eq:power},  ensures
\begin{align}
	\sum\limits_{\ell=2}^L\mathbb{E}[(X^{(H)}_{\ell}-Y^{(H)}_{\ell})^2]=\sum\limits_{\ell=2}^L\mathbb{E}[(X^{(S)}_{\ell}-Y^{(S)}_{\ell})^2].\label{eq:HS}
\end{align}
By \eqref{eq:total} and \eqref{eq:alpha},
\begin{align}
	I(X^{(S)}_1;W^{(S)}_1)&=\frac{1}{2}\log(\Gamma+1)-\sum\limits_{\ell=2}^LI(X^{(S)}_{\ell};W^{(S)}_{\ell})\nonumber\\
	&=\frac{1}{2}\log((1-\alpha)\Gamma+1).
\end{align}
This, together with the fact that
\begin{align}
	I(X^{(S)}_1;W^{(S)}_1)=\frac{1}{2}\log\left(\frac{\lambda_1}{\delta^{(S)}_1(\Gamma)}\right),
\end{align}
implies
\begin{align}
	\delta^{(S)}_1(\Gamma)=\frac{\lambda_1}{(1-\alpha)\Gamma+1}.
\end{align}
Therefore, 
\begin{align}
	\mathbb{E}[(X^{(S)}_1-Y^{(S)}_1)^2]=2\delta^{(S)}_1(\Gamma)=\frac{2\lambda_1}{(1-\alpha)\Gamma+1}.\label{eq:S}
\end{align}
In contrast,
\begin{align}
	\mathbb{E}[(X^{(H)}_1-Y^{(H)}_1)^2]
	&=\mathbb{E}\left[\left(X^{(H)}_1-\sqrt{\frac{\lambda_1}{(1-\alpha)\Gamma+1}}\left(\sqrt{\frac{(1-\alpha)\Gamma}{\lambda_1}}X^{(H)}_1+N\right)\right)^2\right]\nonumber\\
	&=2\left(1-\sqrt{\frac{(1-\alpha)\Gamma}{(1-\alpha)\Gamma+1}}\right)\lambda_1.\label{eq:H}
\end{align}
Combining \eqref{eq:HS}, \eqref{eq:S}, and \eqref{eq:H} shows
\begin{align}
	\mathbb{E}[\|X^{(H)}-Y^{(H)}\|^2]-\mathbb{E}[\|X^{(S)}-Y^{(S)}\|^2]&=\mathbb{E}[(X^{(H)}_1-Y^{(H)}_1)^2]-\mathbb{E}[(X^{(S)}_1-Y^{(S)}_1)^2]\nonumber\\
	&=2\left(\frac{(1-\alpha)\Gamma}{(1-\alpha)\Gamma+1}-\sqrt{\frac{(1-\alpha)\Gamma}{(1-\alpha)\Gamma+1}}\right)\lambda_1\nonumber\\
	&<0
\end{align}
for $\Gamma>0$, confirming the superiority  of the hybrid coding scheme as claimed.
\end{enumerate}

\begin{figure}[htbp]
	\centerline{\includegraphics[width=12cm]{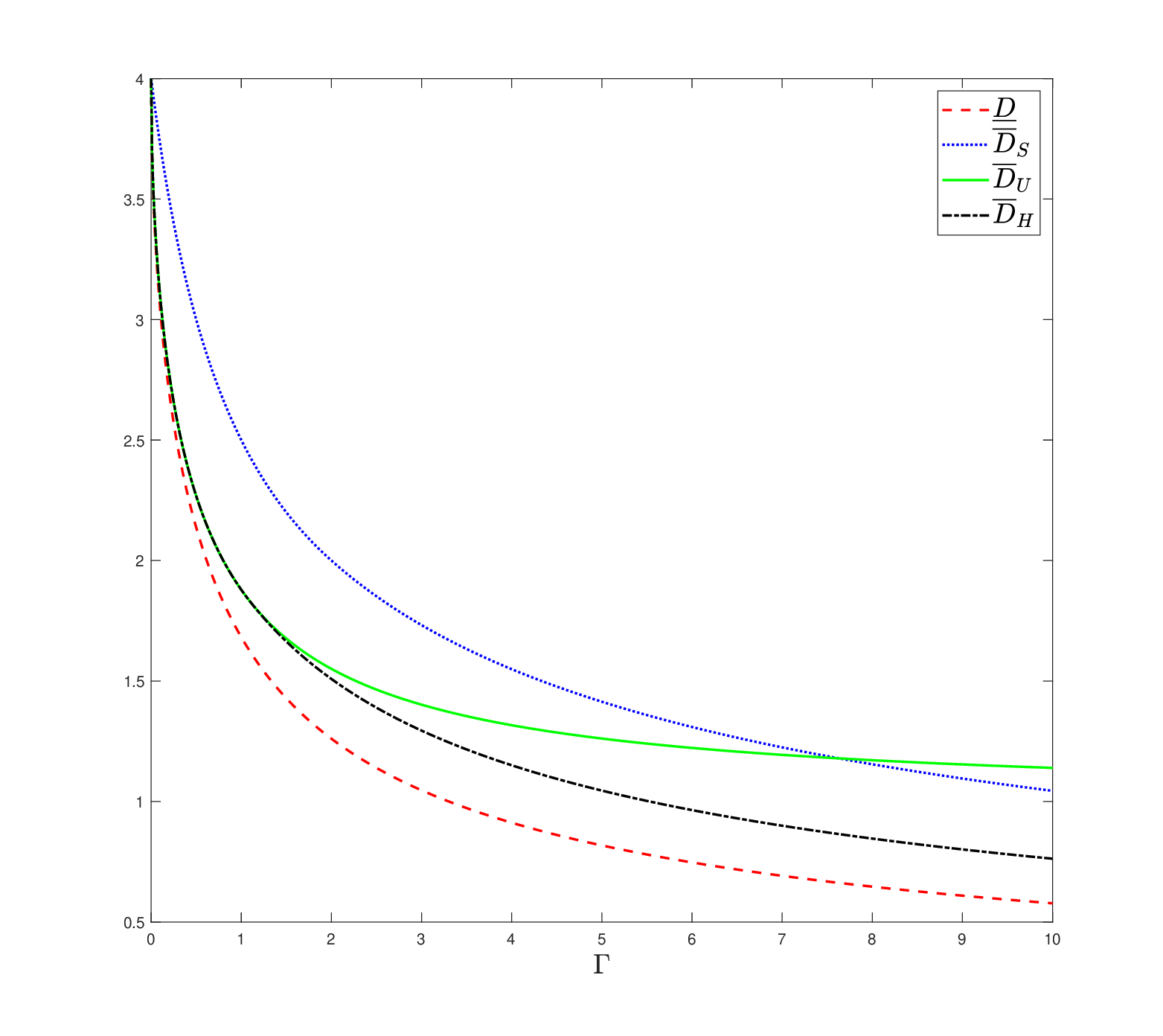}} \caption{Plots of $\underline{D}(\Gamma)$, $\overline{D}_S(\Gamma)$, $\overline{D}_U(\Gamma)$,  and $\overline{D}_H(\Gamma)$  for the Gaussian case with $\Sigma=\mathrm{diag}(\frac{3}{2},\frac{1}{2})$.}
	\label{fig:Gaussian} 
\end{figure}

In Fig. \ref{fig:Gaussian}, we plot $\underline{D}(\Gamma)$, $\overline{D}_S(\Gamma)$, $\overline{D}_U(\Gamma)$, and $\overline{D}_H(\Gamma)$ for the special case $\Sigma=\mathrm{diag}(\frac{3}{2},\frac{1}{2})$. It can be seen that the uncoded scheme outperforms the separation-based scheme in the lower-power regime, but falls behind in the high-power regime. Both schemes, however, are dominated by the hybrid coding scheme. Nonetheless, the hybrid coding scheme still fails to attain $\underline{D}(\Gamma)$---the minimum achievable distortion in the scenario where the encoder and decoder have access to unlimited common randomness.

\begin{figure}[htbp]
	\centerline{\includegraphics[width=12cm]{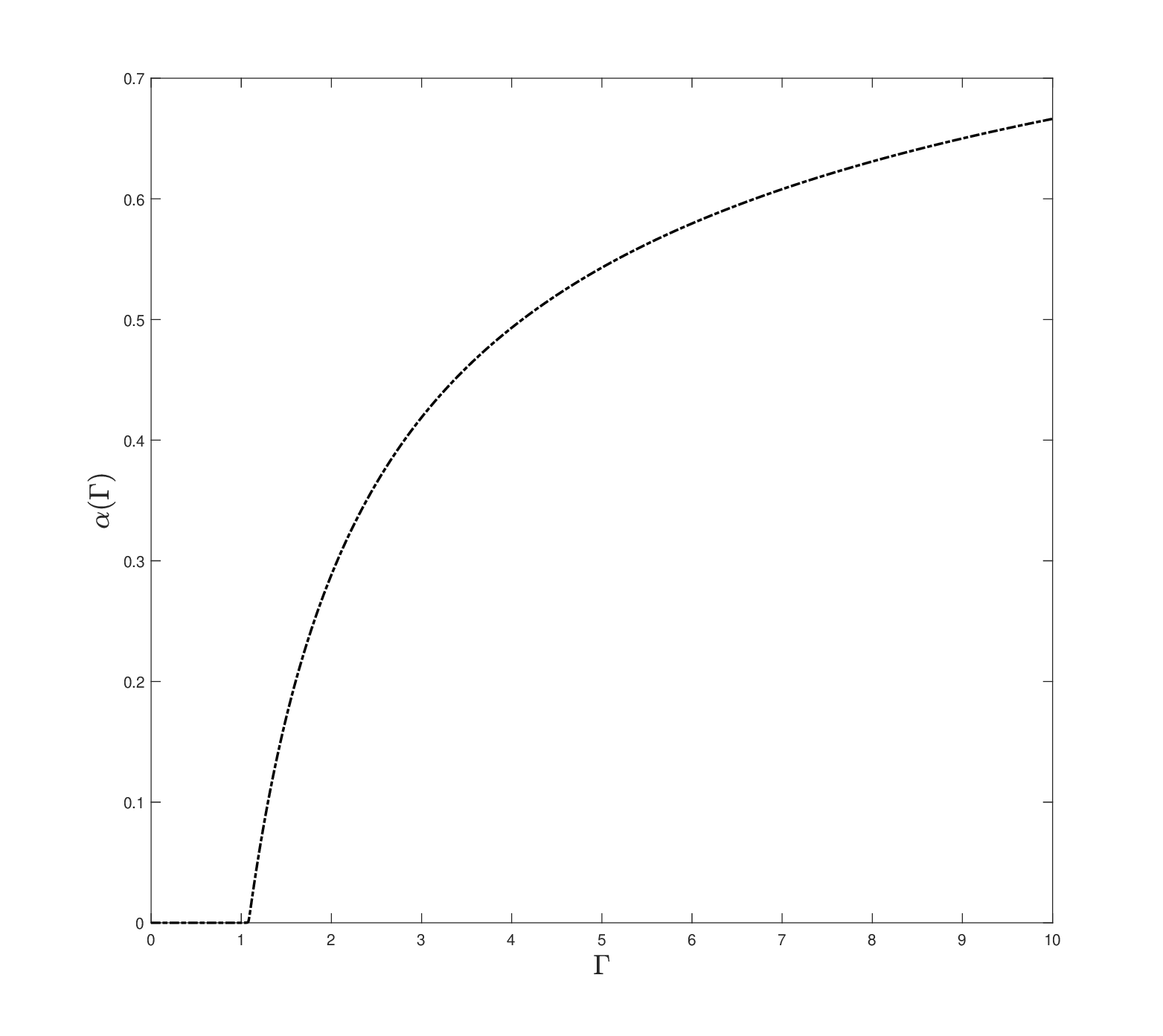}} \caption{Plot of $\alpha(\Gamma)$ for the Gaussian case with $\Sigma=\mathrm{diag}(\frac{3}{2},\frac{1}{2})$.}
	\label{fig:optimal_alpha} 
\end{figure}

Fig. \ref{fig:Gaussian} indicates that $\overline{D}_H(\Gamma)$ coincides with $\overline{D}_U(\Gamma)$ when $\Gamma$ is below a certain threshold $\Gamma^*$. To confirm this, we plot the minimizer of \eqref{eq:minimizer}, denoted by $\alpha(\Gamma)$, against $\Gamma$ in Fig. \ref{fig:optimal_alpha}. It can be seen that for $\Gamma\leq\Gamma^*$,  indeed  $\alpha(\Gamma)=0$, meaning all the power is allocated to the analog part  and consequently the hybrid coding scheme reduces to the uncoded scheme. The analytical expression of $\Gamma^*$ can be determined as follows. 
Let $L_2:=\max\{\ell\in\{2,\ldots,L\}:\lambda_{\ell}=\lambda_2\}$. When\footnote{We set $\lambda_{L_2+1}:=0$ if $L_2=L$.}
\begin{align}
	\left(\frac{(1-\alpha)\Gamma+1}{\Gamma+1}\right)^{\frac{1}{L_2-1}}\lambda_2\geq\lambda_{L_2+1},\label{eq:L2}
\end{align}
we have
\begin{align}
	\omega_H(\Gamma,\alpha)=\left(\frac{(1-\alpha)\Gamma+1}{\Gamma+1}\right)^{\frac{1}{L_2-1}}\lambda_2
\end{align}
and consequently
\begin{align}
	\overline{D}_H(\Gamma,\alpha)=2\left(\left(1-\sqrt{\frac{(1-\alpha)\Gamma}{(1-\alpha)\Gamma+1}}\right)\lambda_1+(L_2-1)\left(\frac{(1-\alpha)\Gamma+1}{\Gamma+1}\right)^{\frac{1}{L_2-1}}\lambda_2\right). \label{eq:DH}
\end{align}
Since condition \eqref{eq:L2} is satisfied for $\alpha$ sufficiently close to zero, it follows by \eqref{eq:DH} that 
\begin{align}
	\left.\frac{\partial\overline{D}_H(\Gamma,\alpha)}{\partial \alpha}\right|_{\alpha=0}=\frac{\sqrt{\Gamma}\lambda_1}{(\Gamma+1)^{\frac{3}{2}}}-\frac{2\Gamma\lambda_2}{\Gamma+1}.\label{eq:threshold}
\end{align}
The threshold $\Gamma^*$ is the solution to 
\begin{align}
	\frac{\sqrt{\Gamma}\lambda_1}{(\Gamma+1)^{\frac{3}{2}}}-\frac{2\Gamma\lambda_2}{\Gamma+1}=0,
\end{align}
which gives
\begin{align}
	\Gamma^*=\frac{-\lambda_2+\sqrt{\lambda^2_1+\lambda^2_2}}{2\lambda_2}.
\end{align}
In particular, $\Gamma^*=\sqrt{\frac{5}{2}}-\frac{1}{2}\approx1.081$ when $\lambda_1=\frac{3}{2}$ and $\lambda_2=\frac{1}{2}$, consistent with what is shown in Fig. \ref{fig:Gaussian} and Fig. \ref{fig:optimal_alpha}.

\section{Conclusion}\label{sec:conclusion}

We have studied the problem of channel-aware optimal transport. Unlike the classical source-channel communication problem, the optimality of the separation-based architecture in this context depends critically on the availability of common randomness.

In the absence of common randomness, a hybrid coding scheme is proposed that can outperform both separation-based and uncoded schemes. Proving a matching converse for this scheme or identifying further enhancements would be a valuable pursuit.

This paper focuses on the bandwidth-matched case, where the lengths of the source and reconstruction sequences are equal to the number of channel uses. While it is straightforward to demonstrate that, with unlimited common randomness, the optimality of the source-channel separation architecture extends to bandwidth-mismatched scenarios, adapting the proposed hybrid coding scheme to cases of bandwidth expansion or compression poses significant challenges. A particularly intriguing direction for future work lies in investigating the one-shot version of channel-aware optimal transport and leveraging the resulting insights to inform the design of generative joint source-channel coding systems.

\appendices

\section{Proof of Theorem \ref{thm:joint_with_common_randomness}}\label{app:Theorem1}

According to Definition \ref{def:OTCw}, for any $\epsilon>0$, there exist seed distribution $p_Q$, encoding distribution $p_{U^n|X^nQ}$, and decoding distribution $p_{Y^n|V^nQ}$
such that 
\begin{align}
	&p_{Y^n}=p^n_Y,\\
	&\frac{1}{n}\sum\limits_{t=1}^n\mathbb{E}[c(U_t)]\leq\Gamma+\epsilon,\\
	&\frac{1}{n}\sum\limits_{t=1}^n\mathbb{E}[d(X_t,Y_t)]\leq \underline{D}_J(\Gamma)+\epsilon.
\end{align}
Let $T$ be uniformly distributed over $\{1,2,\ldots,n\}$ and independent of $(X^n,Y^n,U^n,V^n,Q)$. 
We have
\begin{align}
	\frac{1}{n}I(X^n;Y^n)&=\frac{1}{n}\sum\limits_{t=1}^nI(X_t;Y^n|X^{t-1})\nonumber\\
	&=\frac{1}{n}\sum\limits_{t=1}^nI(X_t;Y^n,X^{t-1})\nonumber\\
	&\geq\frac{1}{n}\sum\limits_{t=1}^nI(X_t;Y_t)\nonumber\\
	&=I(X_T;Y_T|T)\nonumber\\
	&=I(X_T;Y_T,T)\nonumber\\
	&\geq I(X_T;Y_T).\label{eq:comb1}
\end{align}
On the other hand,
\begin{align}
	\frac{1}{n}I(X^n;Y^n)&\leq \frac{1}{n}I(X^n;Y^n,Q)\nonumber\\
	&=\frac{1}{n}I(X^n;Y^n|Q)\nonumber\\
	&\leq\frac{1}{n}I(U^n;V^n|Q)\nonumber\\
	&=\frac{1}{n}\sum\limits_{t=1}^nI(U^n;V_t|V^{t-1},Q)\nonumber\\
	&=I(U^n;V_T|V^{T-1},Q,T)\nonumber\\
	&\leq I(U^n,V^{T-1},Q,T;V_T)\nonumber\\
	&=I(U_T;V_T).\label{eq:comb2}
\end{align}
Combining \eqref{eq:comb1} and \eqref{eq:comb2} gives
\begin{align}
	I(X_T;Y_T)\leq I(U_T;V_T).\label{eq:XYleqUV}
\end{align}
Since $p_{V_T|U_T}=p_{V|U}$ and
\begin{align}
	\mathbb{E}[c(U_T)]&=\mathbb{E}[\mathbb{E}[c(U_T)|T]]\nonumber\\
	&=\frac{1}{n}\sum\limits_{t=1}^n\mathbb{E}[c(U_t)]\nonumber\\
	&\leq\Gamma+\epsilon,
\end{align}
it follows that 
\begin{align}
	I(U_T;V_T)\leq C(p_{V|U},\Gamma+\epsilon).\label{eq:cost}
\end{align}
Substituting \eqref{eq:cost} into \eqref{eq:XYleqUV} yields
\begin{align}
	I(X_T;Y_T)\leq C(p_{V|U},\Gamma+\epsilon),
\end{align}	
which, together with the fact $p_{X_T}=p_X$
and $p_{Y_T}=p_Y$, further implies
\begin{align}
	\mathbb{E}[d(X_T,Y_T)]\geq \underline{D}_S(\Gamma+\epsilon).\label{eq:comb3}
\end{align}
Moreover, it can be verified that
\begin{align}
	\mathbb{E}[d(X_T,Y_T)]&=\mathbb{E}[\mathbb{E}[d(X_T,Y_T)|T]]\nonumber\\
	&=\frac{1}{n}\sum\limits_{t=1}^n\mathbb{E}[d(X_t,Y_t)]\nonumber\\
	&\leq\underline{D}_J(\Gamma)+\epsilon.\label{eq:comb4} 
\end{align}
Therefore,	
\begin{align}
	\underline{D}_J(\Gamma)+\epsilon\geq\underline{D}_S(\Gamma+\epsilon).
\end{align}	
Sending $\epsilon\rightarrow 0$ and invoking the standard continuity argument proves the desired result.	

\section{Proof of Theorem \ref{thm:hybrid}}\label{app:hybrid}

First consider the case where $\max\{I(X;Z),I(Y;Z)\}<I(Z;V)$. We choose $R$ such that $\max\{I(X;Z),I(Y;Z)\}<R<I(Z;V)$ and develop a hybrid coding scheme based on the given joint distribution $p_{X}p_{ZU|X}p_{V|U}p_{Y|ZV}$ as follows.
\begin{itemize}
	\item Codebook generation: Construct a random
	codebook $\mathcal{C}:=\{Z^{n}(m):m\in\{1,\ldots,\lceil 2^{nR}\rceil\}\}$, where $Z^{n}(m)$, $m=1,\ldots,\lceil 2^{nR}\rceil\}$,
	are generated independently according to  $p_{Z}^{n}$.
	
	\item Encoding: Stochastically map the source sequence $X^{n}$ to a message $M$ using the likelihood encoder \cite{SCP16}
	\begin{align}
		p_{M|X^{n}\mathcal{C}}(m|x^{n},\mathcal{C}):=\frac{p_{X|Z}^{n}(x^{n}|Z^{n}(m))}{\sum_{m=1}^{\lceil 2^{nR}\rceil}p_{X|Z}^{n}(x^{n}|Z^{n}(m))}.
	\end{align}
	and generate
	$U^{n}$ based on $(X^{n},Z^{n}(M))$ in a symbol-by-symbol manner
	through $p_{U|XZ}$ (we need $R>I(X;Z)$ to ensure successful encoding).
	
	\item Decoding: Given the channel output $V^{n}$, produce a decoded message $\hat{M}$
	using the joint typicality decoding rule (we need $R> I(Z;V)$
	to ensure successful decoding). Then generate $\hat{Y}^{n}$ based on $(Z^{n}(\hat{M}),V^{n})$
	in a symbol-by-symbol manner through $p_{Y|ZV}$ such that $\hat{Y}^{n}$ approximately
	follows $p_{Y}^{n}$ (we need $R> I(Y;Z)$ to ensure soft covering).
	Finally, generate the reconstruction sequence $Y^{n}$ based on $\hat{Y}^{n}$ through $p_{Y^{n}|\hat{Y}^{n}\mathcal{C}}$ induced by a maximal
	coupling $p_{\hat{Y}^{n}Y^n|\mathcal{C}}$ of $p_{\hat{Y}^{n}|\mathcal{C}}$ and $p_{Y}^{n}$.
\end{itemize}

We now procced to give a detailed performance analysis. Let $p_{X^{n}MU^{n}V^{n}\hat{M}\hat{Y}^{n}Y^{n}|\mathcal{C}}$ be the joint distribution induced by this scheme given the random codebook $\mathcal{C}$: 
\begin{align*}
	& p_{X^{n}MU^{n}V^{n}\hat{M}\hat{Y}^{n}Y^{n}|\mathcal{C}}(x^{n},m,u^{n},v^{n},\hat{m},\hat{y}^{n},y^{n}|\mathcal{C})\\
	& :=p_{X}^{n}(x^{n})p_{M|X^{n}\mathcal{C}}(m|x^{n},\mathcal{C})p_{U|XZ}^{n}(u^{n}|x^{n},Z^{n}(m))p_{V|U}^{n}(v^{n}|u^{n}) p_{\hat{m}|V^{n}\mathcal{C}}(\hat{m}|v^{n},\mathcal{C})p_{Y|ZV}^{n}(\hat{y}^{n}|Z^{n}(\hat{m}),v^{n})p_{Y^{n}|\hat{Y}^{n}\mathcal{C}}(y^{n}|\hat{y}^{n},\mathcal{C}).
\end{align*}
Define auxiliary joint distributions 
\begin{align}
	& \bar{p}_{X^{n}MU^{n}V^{n}\hat{M}\hat{Y}^{n}Y^{n}|\mathcal{C}}(x^{n},m,u^{n},v^{n},\hat{m},\hat{y}^{n},y^{n}|\mathcal{C})\nonumber\\
	& :=\bar{p}_{M}(m)p_{X|Z}^{n}(x^{n}|Z^{n}(m))p_{U|XZ}^{n}(u^{n}|x^{n},Z^{n}(m))p_{V|U}^{n}(v^{n}|u^{n}) p_{\hat{m}|V^{n}\mathcal{C}}(\hat{m}|v^{n},\mathcal{C})p_{Y|ZV}^{n}(\hat{y}^{n}|Z^{n}(\hat{m}),v^{n})p_{Y^{n}|\hat{Y}^{n}\mathcal{C}}(y^{n}|\hat{y}^{n},\mathcal{C})
\end{align}
and 
\begin{align}
	& \bar{q}_{X^{n}MU^{n}V^{n}\hat{M}\hat{Y}^{n}Y^{n}|\mathcal{C}}(x^{n},m,u^{n},v^{n},\hat{m},\hat{y}^{n},y^{n}|\mathcal{C})\nonumber\\
	& :=\bar{p}_{M}(m)p_{X|Z}^{n}(x^{n}|Z^{n}(m))p_{U|XZ}^{n}(u^{n}|x^{n},Z^{n}(m))p_{V|U}^{n}(v^{n}|u^{n}) q_{\hat{M}|M}(\hat{m}|m)p_{Y|ZV}^{n}(\hat{y}^{n}|Z^{n}(\hat{m}),v^{n})p_{Y^{n}|\hat{Y}^{n}\mathcal{C}}(y^{n}|\hat{y}^{n},\mathcal{C}),
\end{align}
where 
\begin{align}
	&\bar{p}_{M}(m)  :=\frac{1}{\lceil 2^{nR}\rceil}1\{m\in\{1,\ldots,\lceil 2^{nR}\rceil\}\},\\
	&q_{\hat{M}|M}(\hat{m}|m) :=1\{\hat{m}=m\}.
\end{align}
Since $p_{M|X^{n}\mathcal{C}}=\bar{p}_{M|X^{n}\mathcal{C}}$,
it follows that
\begin{align}
	d_{TV}(p_{X^{n}MU^{n}V^{n}\hat{M}\hat{Y}^{n}Y^{n}|\mathcal{C}},\bar{p}_{X^{n}MU^{n}V^{n}\hat{M}\hat{Y}^{n}Y^{n}|\mathcal{C}}) & =d_{TV}(p_{X}^{n},\bar{p}_{X^{n}|\mathcal{C}}),\label{eq:TV}
\end{align}
where $d_{TV}(\cdot,\cdot)$ denotes the total variation distance. As $R>I(X;Z)$, by the soft-covering lemma \cite[Theorem 1]{Cuff2016},  given any $\delta>0$, we have\footnote{We use ``w.h.p." as an abbreviation for ``with high probability", which means, throughout this proof, that the probability (with respect to the randomly generated codebook $\mathcal{C}$) approaches one as $n\rightarrow\infty$.}
\begin{align}
	d_{TV}(p_{X}^{n},\bar{p}_{X^{n}|\mathcal{C}})\le\delta\quad\mbox{w.h.p.},
\end{align}
which, together with \eqref{eq:TV}, implies
\begin{align}
	& d_{TV}(p_{X^{n}MU^{n}V^{n}\hat{M}\hat{Y}^{n}Y^{n}|\mathcal{C}},\bar{p}_{X^{n}MU^{n}V^{n}\hat{M}\hat{Y}^{n}Y^{n}|\mathcal{C}})\le\delta\quad\mbox{w.h.p.}\label{eq:-1}
\end{align}
In light of  \cite[Lemma 2]{SCP16}, 
\begin{align}
	d_{TV}(\bar{p}_{X^{n}MU^{n}V^{n}\hat{M}\hat{Y}^{n}Y^{n}|\mathcal{C}},\bar{q}_{X^{n}MU^{n}V^{n}\hat{M}\hat{Y}^{n}Y^{n}|\mathcal{C}}) & \le\bar{p}\{\hat{M}\neq M|\mathcal{C}\}.
\end{align}
Moreover, under the distribution $\bar{p}$, the probability of error induced by joint typicality decoding satisfies
\begin{align}
	\bar{p}\{\hat{M}\neq M|\mathcal{C}\}\le\delta\quad\mbox{w.h.p.}
\end{align}
Therefore, 
\begin{align}
	d_{TV}(\bar{p}_{X^{n}MU^{n}V^{n}\hat{M}\hat{Y}^{n}Y^{n}|\mathcal{C}},\bar{q}_{X^{n}MU^{n}V^{n}\hat{M}\hat{Y}^{n}Y^{n}|\mathcal{C}})\le\delta\quad\mbox{w.h.p.}\label{eq:-2}
\end{align}
Combining (\ref{eq:-1}) and (\ref{eq:-2}) shows
\begin{align}
	d_{TV}(p_{X^{n}MU^{n}V^{n}\hat{M}\hat{Y}^{n}Y^{n}|\mathcal{C}},\bar{q}_{X^{n}MU^{n}V^{n}\hat{M}\hat{Y}^{n}Y^{n}|\mathcal{C}})\le2\delta\quad\mbox{w.h.p.}\label{eq:-2-1}
\end{align}
Note that
\begin{align*}
	& \bar{q}_{X^{n}MU^{n}V^{n}\hat{Y}^{n}Y^{n}|\mathcal{C}}(x^{n},m,u^{n},v^{n},\hat{y}^{n},y^{n}|\mathcal{C})\\
	& =\bar{p}_{M}(m)p_{X|Z}^{n}(x^{n}|Z^{n}(m))p_{U|XZ}^{n}(u^{n}|x^{n},Z^{n}(m))p_{V|U}^{n}(v^{n}|u^{n})p_{Y|ZV}^{n}(\hat{y}^{n}|Z^{n}(m),v^{n})p_{Y^{n}|\hat{Y}^{n}}(y^{n}|\hat{y}^{n},\mathcal{C}).
\end{align*}
Since
\begin{align}
	\bar{q}\{Z^n(M)\in\mathcal{T}^{(n)}_{\delta}(p_Z)|\mathcal{C}\}\geq1-\delta\quad\mbox{w.h.p.},
\end{align}
it follows by the conditional typicality lemma \cite[p. 27]{EGK11}  that
\begin{align*}
	\bar{q}\{(Z^n(M),X^n,U^n,\hat{Y}^n)\in\mathcal{T}^{(n)}_{2\delta}(p_{ZXUY})|\mathcal{C}\}\geq 1-2\delta\quad\mbox{w.h.p.},	
\end{align*} 
which, in light of the typical average lemma \cite[p. 26]{EGK11}, implies
\begin{align*}
	&\frac{1}{n}\sum\limits_{t=1}^n\mathbb{E}_{\bar{q}}[c(U_t)|\mathcal{C}]  \le(1+2\delta)\mathbb{E}_{p}[c(U)]+2\delta c_{\max}\quad\mbox{w.h.p.},\\
	&\frac{1}{n}\sum\limits_{t=1}^n\mathbb{E}_{\bar{q}}[d(X_t,\hat{Y}_t)|\mathcal{C}]  \le(1+2\delta)\mathbb{E}_{p}[d(X,Y)]+2\delta d_{\max}\quad\mbox{w.h.p.},
\end{align*}
where $c_{\max}:=\max_{u\in\mathcal{U}}c(u)$ and $d_{\max}:=\max_{x\in\mathcal{X},y\in\mathcal{Y}}d(x,y)$.
By (\ref{eq:-2-1}), 
\begin{align}
	&\frac{1}{n}\sum\limits_{t=1}^n\mathbb{E}_{p}[c(U_t)|\mathcal{C}]  \le(1+2\delta)\mathbb{E}_{p}[c(U)]+4\delta c_{\max},\label{eq:-5}\\
	&\frac{1}{n}\sum\limits_{t=1}^n\mathbb{E}_{p}[d(X_t,\hat{Y}_t)|\mathcal{C}] \le(1+2\delta)\mathbb{E}_{p}[d(X,Y)]+4\delta d_{\max}.\label{eq:-4}
\end{align}
Since $p_{\hat{Y}^{n}Y^{n}|\mathcal{C}}$ is a maximal coupling of $p_{\hat{Y}^{n}|\mathcal{C}}$
and $p_{Y}^{n}$, we have
\begin{equation}
	p\{\hat{Y}^{n}\neq Y^{n}|\mathcal{C}\}=d_{TV}(p_{Y^{n}|\mathcal{C}}, p_{Y}^{n}).\label{eq:inview1}
\end{equation}
As $R>I(Y;Z)$, by the soft-covering lemma \cite[Theorem 1]{Cuff2016}, 
\begin{align}
	d_{TV}(\bar{q}_{Y^{n}|\mathcal{C}}, p_{Y}^{n})\le\delta\quad\mbox{w.h.p.}\label{eq:inview2}
\end{align}
In view of \eqref{eq:-2-1}, \eqref{eq:inview1}, and \eqref{eq:inview2},
\begin{equation}
	p\{\hat{Y}^{n}\neq Y^{n}|\mathcal{C}\}\le d_{TV}(p_{Y^{n}|\mathcal{C}},\bar{q}_{Y^{n}|\mathcal{C}})+d_{TV}(\bar{q}_{Y^{n}|\mathcal{C}},p_{Y}^{n})\le3\delta,\label{eq:-3}
\end{equation}
which, together with (\ref{eq:-4}), implies 
\begin{equation}
	\frac{1}{n}\sum\limits_{t=1}^n\mathbb{E}_{p}[d(X_t,Y_t)|\mathcal{C}]\le(1+2\delta)\mathbb{E}_{p}[d(X,Y)]+7\delta d_{\max}\quad\mbox{w.h.p.}\label{eq:-6}
\end{equation}
Invoking (\ref{eq:-5}), (\ref{eq:-6}), and the fact that  $\delta>0$ is arbitrary
establishes the desired achievability result.


Next consider the case where $\max\{I(X;Z),I(Y;Z)\}=I(Z;V)$. If $I(Z;V)=0$, then $I(X;Z)=I(Y;Z)=0$, i.e.,  
$Z$ is pairwise independent of $X$ and $Y$. As a result, the role of $Z$ is simply to direct time-sharing among several uncoded schemes, each associated with a specific realization of $Z$. 
If $I(Z;V)>0$, then it is possible construct a sequence of augmented joint distributions  $\{p_{X^{(k)}XZUVYY^{(k)}}:=p_{X^{(k)}|X}p_{XZUVY}p_{Y^{(k)}|Y}\}_{k=1}^{\infty}$ such that
\begin{enumerate}
	\item $p_{X^{(k)}}=p_X$ and $p_{Y^{(k)}}=p_Y$ for all $k$,
	
	\item $\max\{I(X^{(k)};Z),I(Y^{(k)};Z)\}<I(Z;V)$ for all $k$,
	
	\item $\lim_{k\rightarrow\infty}p\{X^{(k)}=X,Y^{(k)}=Y\}=1$.
\end{enumerate}
For example, we can let 
\begin{align}
	&p_{X^{(k)}|X}(x'|x):=\begin{cases}
		\frac{(k+1)p_X(x)-p^{\min}_X}{(k+1)p(x)}, & x=x',\\
		\frac{p^{\min}_{X}}{(k+1)(|\mathcal{X}|-1)p_X(x)}, & x\neq x',
	\end{cases}\\
	&p_{Y^{(k)}|Y}(y'|y):=\begin{cases}
		\frac{(k+1)p_Y(y)-p^{\min}_Y}{(k+1)p(y)}, & y=y',\\
		\frac{p^{\min}_{Y}}{(k+1)(|\mathcal{Y}|-1)p_Y(y)}, & y\neq y',
	\end{cases}
\end{align}
where $p^{\min}_{X}:=\min_{x\in\mathcal{X}}p_X(x)$ and $p^{\min}_{Y}:=\min_{y\in\mathcal{Y}}p_Y(y)$. Clearly, with this construction, properties 1) and 2) are satisfied. Moreover, since $p_{XX^{(k)}}$ is indecomposable, it follows by \cite[p. 402, Problem 25]{CK81} that $I(X^{(k)};Z)<I(X;Z)$ or $I(X^{(k)};Z)=0$. In either case, we have $I(X^{(k)};Z)<I(Z;V)$. Similarly,  $I(Y^{(k)};Z)<I(Z;V)$. Hence, property 3) is also satisfied.
In view of properties 1) and 2), the preceding argument can be readily leveraged to show 
$\mathbb{E}[d(X^{(k)},Y^{(k)})]$ is achievable. It then follows by property 3) that 
\begin{align}
	\mathbb{E}[d(X,Y)]=\lim\limits_{k\rightarrow\infty}\mathbb{E}[d(X^{(k)},Y^{(k)})]
\end{align}
and is therefore also achievable.

Finally, the cardinality bound $|\mathcal{Z}|\leq|\mathcal{X}|+|\mathcal{Y}|+|\mathcal{V}|+2$ follows by the observation that it suffices to preserve $p_X$, $p_Y$, $p_V$, $H(X|Z)$, $H(Y|Z)$, $H(V|Z)$, $\mathbb{E}[c(U)]$, and $\mathbb{E}[d(X,Y)]$.

\section{On the Optimal Linear Scheme}\label{app:linear}

\begin{theorem}
	Let $X\sim\mathcal{N}(0,\Sigma)$ and $N\sim\mathcal{N}(0,1)$ be mutually independent, where $\Sigma:=\mathrm{diag}(\lambda_1,\ldots,\lambda_L)$ with $\lambda_1\geq\ldots\geq\lambda_L>0$. For any $g:=(g_1,\ldots,g_L)^T\in\mathbb{R}^L$ and $Y\sim\mathcal{N}(0,\Lambda)$ such that $X\leftrightarrow (g^TX+N)\leftrightarrow Y$ form a Markov chain, we have
	\begin{align}
		\mathbb{E}[\|X-Y\|^2]\geq\overline{D}_U(g^T\Sigma g).\label{eq:uncoded}
	\end{align}
\end{theorem}
\begin{IEEEproof}
	It is straightforward see that \eqref{eq:uncoded} holds with equality when $\|g\|=0$. Henceforth, we shall assume $\|g\|>0$.
	
	Note that 
	\begin{align}
		\mathbb{E}[X|g^TX+N]=s^T\tilde{X},
	\end{align}
	where $s:=\frac{1}{\|\Sigma g\|}\Sigma g$ and $\tilde{X}:=\frac{\|\Sigma g\|}{g^T\Sigma g+1}(g^TX+N)$. Since $X\leftrightarrow (g^TX+N)\leftrightarrow Y$ form a Markov chain,
	\begin{align}
		\mathbb{E}[\|X-Y\|^2]=\mathbb{E}[\|X-s^T\tilde{X}\|^2]+\mathbb{E}[\|Y-s^T\tilde{X}\|^2].\label{eq:desired}
	\end{align}
	We now proceed to derive lower bounds for $\mathbb{E}[\|X-s^T\tilde{X}\|^2]$ and $\mathbb{E}[\|Y-s^T\tilde{X}\|^2]$ separately.
	
	\begin{itemize}
		\item We have
		\begin{align}
			\mathbb{E}[\|X-s^T\tilde{X}\|^2]=\mathrm{tr}(\Sigma-\gamma ss^T)=\sum\limits_{\ell=1}^L\lambda_{\ell}-\gamma,\label{eq:lowerbound1}
		\end{align}
		where $\gamma:=\frac{\|\Sigma g\|^2}{g^T\Sigma g+1}$. Note that
		\begin{align*}
			I(g^TX;g^TX+N)=I(X;g^TX+N)=I(X;s^T\tilde{X}),
		\end{align*}
		which, together with the fact
		\begin{align}
			&I(g^TX;g^TX+N)=\frac{1}{2}\log(g^T\Sigma g+1),\\
			&I(X;s^T\tilde{X})=\frac{1}{2}\log\left(\frac{|\Sigma|}{|\Sigma-\gamma ss^T|}\right),
		\end{align}
		implies
		\begin{align}
			g^T\Sigma g+1=\frac{|\Sigma|}{|\Sigma-\gamma ss^T|}.\label{eq:twoway}
		\end{align}
		By the Sherman-Morrison formula,
		\begin{align}
			\frac{|\Sigma|}{|\Sigma-\gamma ss^T|}=\frac{1}{1-\gamma\sum_{\ell=1}^L\frac{\lambda_{\ell}g^2_{\ell}}{\|\Sigma g\|^2}}.
		\end{align}
		Since $\sum_{\ell=1}^L\frac{\lambda_{\ell}g^2_{\ell}}{\|\Sigma g\|^2}\geq\frac{1}{\lambda_1}$, it follows that
		\begin{align}
			\frac{|\Sigma|}{|\Sigma-\gamma ss^T|}\geq\frac{1}{1-\frac{\gamma}{\lambda_1}},
		\end{align}
		which, in conjunction with \eqref{eq:twoway}, further implies
		\begin{align}
			\gamma\leq\frac{g^T\Sigma g\lambda_1}{g^T\Sigma g+1}.\label{eq:sub1}
		\end{align}
		Substituting \eqref{eq:sub1} into \eqref{eq:lowerbound1} gives
		\begin{align}
			\mathbb{E}[\|X-s^T\tilde{X}\|^2]\geq\frac{\lambda_1}{g^T\Sigma g+1}+\sum\limits_{\ell=2}^L\lambda_{\ell}.\label{eq:item1}
		\end{align}

		\item We have
		\begin{align}
			\mathbb{E}[\|Y-s_1^T\tilde{X}\|^2]\geq W^2_2(\mathcal{N}(0,\Sigma),\mathcal{N}(0,\gamma s_1s^T_1)),\label{eq:term2}
		\end{align}
		where $W_2(\cdot,\cdot)$ denotes the Wasserstein-$2$ distance. Let $\Pi$ be a unitary matrix with the first column being $s$. It can be verified that
		\begin{align}
			W^2_2(\mathcal{N}(0,\Sigma),\mathcal{N}(0,\gamma s_1s^T_1))&=W^2_2(\mathcal{N}(0,\Pi^T\Sigma \Pi),\mathcal{N}(0,\mathrm{diag}(\gamma,0,\ldots,0)))\nonumber\\
			&=\left(\sqrt{\sum\limits_{\ell=1}^L\frac{\lambda^3_{\ell}g^2_{\ell}}{\|\Sigma g\|^2}}-\sqrt{\gamma}\right)^2+\mathrm{tr}(\Pi^T\Sigma\Pi)-\sum\limits_{\ell=1}^L\frac{\lambda^3_{\ell}g^2_{\ell}}{\|\Sigma g\|^2}\nonumber\\
			&=\lambda_1-2\sqrt{\gamma\sum\limits_{\ell=1}^L\frac{\lambda^3_{\ell}g^2_{\ell}}{\|\Sigma g\|^2}}+\gamma+\sum\limits_{\ell=2}^L\lambda_{\ell},
		\end{align}
	which, together with the fact $\sum_{\ell=1}^L\frac{\lambda^3_{\ell}g^2_{\ell}}{\|\Sigma g\|^2}\leq\lambda_1$, implies
	\begin{align}
		W^2_2(\mathcal{N}(0,\Sigma),\mathcal{N}(0,\gamma s_1s^T_1))\geq(\sqrt{\lambda_1}-\sqrt{\gamma})^2+\sum\limits_{\ell=2}^L\lambda_{\ell}.\label{eq:W2}
	\end{align}
	 Combining  \eqref{eq:sub1}, \eqref{eq:term2}, and \eqref{eq:W2} shows
		\begin{align}
			\mathbb{E}[\|Y-s_1^T\tilde{X}\|^2]\geq\left(\frac{g^T\Sigma g}{g^T\Sigma g+1}-2\sqrt{\frac{g^T\Sigma g}{g^T\Sigma g+1}}\right)\lambda_1+\sum\limits_{\ell=1}^L\lambda_{\ell}.\label{eq:item2}
		\end{align}
	\end{itemize}

	Substituting \eqref{eq:item1} and \eqref{eq:item2} into \eqref{eq:desired} yields the desired result.	
\end{IEEEproof}

\end{document}